\documentclass{Nature}
\usepackage[normalem]{ulem}
\usepackage{graphicx}
\usepackage{amssymb}
\usepackage{amsmath}
\usepackage{graphicx}
\usepackage{dcolumn,}
\usepackage{bm}
\usepackage{hyperref}
\usepackage{siunitx}
\usepackage{graphicx}
\usepackage{enumitem}
\usepackage{paralist}
\usepackage{xcolor}
\usepackage{xspace}
\usepackage[capitalize]{cleveref}
\usepackage{multirow}
\usepackage{soul}

\newcommand{\be}{\begin{equation}}
\newcommand{\ee}{\end{equation}}
\newcommand{\ba}{\begin{align}}
\newcommand{\ea}{\end{align}}
\newcommand{\ord}{\mathcal{O}}

\newcommand{\chiperp}{\chi_\perp}
\newcommand{\chiPerp}{x}
\newcommand{\chip}{\chi_\text{p}}

\newcommand{\chieff}{\chi_{\text{eff}}}
\newcommand{\chipar}{\chi_{\parallel}}
\newcommand{\chiPar}{y}
\newcommand{\Mtot}{M}
\newcommand{\fit}{\mathcal{F}}
\newcommand{\model}{\mathfrak{M}}
\newcommand{\mismatch}{\mathcal{M}}

\newcommand{\modelname}[1]{\textsc{#1}}

\def\citea#1{\cite{#1}}
\def\citea#1{}
\def\Msun{M_\odot}

\def\f{\frac}
\def\NRsur{{\modelname{NRSur7dq4}}}

\title{Incorporation of model accuracy in gravitational-wave Bayesian inference}

\author{Charlie Hoy$^{1\dagger}$, Sarp Ak\c{c}ay$^{2}$, Jake Mac Uilliam$^{2}$ and Jonathan E. Thompson$^{3,4}$}

\begin{document}

\maketitle

\begin{affiliations}
  \item {Institute of Cosmology and Gravitation, University of Portsmouth, Portsmouth, PO1 3FX, UK}
  \item {School of Mathematics \& Statistics, University College Dublin, Dublin, D4, Ireland}
  \item{Theoretical Astrophysics Group, California Institute of Technology, Pasadena, CA, 91125, U.S.A}
  \item{Mathematical Sciences \& STAG Research Centre, University of Southampton, Southampton, SO17 1BJ, UK. $\dagger$ e-mail: \href{mailto:charlie.hoy@port.ac.uk}{charlie.hoy@port.ac.uk}}
\end{affiliations}

\abstract{
  Inferring the properties of colliding black holes from gravitational-wave observations is subject to systematic errors arising from modelling uncertainties. Although the accuracy of each model can be calculated through comparison to theoretical expectations from general relativity, Bayesian analyses are yet to incorporate this information. As such, a mixture model is typically used where results obtained with different gravitational-wave models are combined with either equal weight, or based on their relative Bayesian evidence. In this work we present a novel method to incorporate the accuracy of multiple models in gravitational-wave Bayesian analyses. By analysing simulated gravitational-wave signals in zero-noise, we show that our technique uses $30\%$ less computational resources, and more faithfully recovers the true parameters than existing techniques. We further apply our method to a real gravitational-wave signal and, when assuming the binary black hole hypothesis, demonstrate that the source of GW191109\_010717 has unequal component masses, with the primary having a $69\%$ probability that it lies above the maximum black hole mass from stellar collapse. We envisage that this method will become an essential tool within ground-based gravitational-wave astronomy.

\maketitle

\normalfont
\setlength{\parindent}{0.20in}

\section{Model systematics in gravitational-wave\\astronomy}
\label{sec:introduction}

Our ability to infer the properties of colliding black holes from an observed gravitational-wave (GW) signal is dependent on our chosen model\cite{Veitch:2014wba}. Models that poorly describe general relativity will not only yield biased results for individual sources\cite{Ramos-Buades:2023ehm,Thompson:2023ase,Yelikar:2024wzm,MacUilliam:2024oif,Dhani:2024jja,Akcay:2025rve}, but also incorrect inferences for the properties of the underlying astrophysical population -- for example, the mass and spin distributions of black holes in the Universe\cite{Purrer:2019jcp,Moore:2021eok,Kapil:2024zdn}. Unbiased results will only be obtained with models that are perfect descriptions of general relativity (assuming a known understanding of the noise in the GW detectors\cite{TheLIGOScientific:2014jea,acernese2014advanced,KAGRA:2020tym}).

Unfortunately, directly computing GW signals from general relativity is a computationally expensive task; numerical relativity simulations, where Einstein's equations of general relativity are solved on high-performance computing clusters, require millions of CPU hours to perform\cite{Hamilton:2023qkv}. For this reason only several thousand simulations are currently available\cite{Mroue:2013xna,Boyle:2019kee,Healy:2017psd, Healy:2019jyf,Healy:2020vre,Healy:2022wdn,Jani:2016wkt,Hamilton:2023qkv}. As a result, {the latest} GW models rely on analytical or semi-analytical prescriptions that are calibrated to the 
numerical relativity simulations\cite{Pratten:2020ceb,Estelles:2021gvs,Thompson:2023ase, Albertini:2021tbt, Nagar:2023zxh,Ramos-Buades:2023ehm,Colleoni:2024knd}, or are based on surrogate modelling techniques\cite{Varma:2019csw, Varma:2018mmi}. However, each modelling approach will incur some degree of approximation errors.

The accuracy of a GW model is typically measured by the mismatch\cite{Owen:1995tm} between the model and a fiducial waveform, often a numerical relativity simulation. The mismatch varies between 0, signifying that the model and the true waveform are identical (up to an overall amplitude rescaling), and 1, meaning that the two are completely orthogonal. It is well known that certain models are more faithful to general relativity than others in different regions of parameter space\cite{MacUilliam:2024oif}.

The standard approach to account for modelling errors when inferring the properties of binary black holes is to construct a mixture model, where results from numerous analyses are combined; a Bayesian analysis is performed for each GW model and the results are either mixed together with equal weights\cite{LIGOScientific:2021djp} or according to their relative Bayesian evidence\cite{Ashton:2019leq}, or by averaging the likelihood\cite{Jan:2020bdz}. An alternative technique involves sampling over a set of GW models in a single {\emph{joint}} Bayesian analysis\cite{Ashton:2021cub,Hoy:2022tst}. Although widely used, these methods do not account for the known accuracy of the GW model. 

Other approaches have suggested quantifying the uncertainty in a GW model and marginalizing over this error in Bayesian analyses\cite{Moore:2014pda,Doctor:2017csx,Williams:2019vub,Read:2023hkv,Khan:2024whs}.  These methods have either not been demonstrated in practice, or are only suitable for a single model. Preliminary work has investigated incorporating model accuracy into likelihood averaging techniques for simplified models\cite{Jan:2021aaa}. However, this approach maintains a comparable computational cost to evidence mixing~\cite{Ashton:2019leq} and is difficult to interface with standard Bayesian inference techniques.

In this work we present the first approach to incorporate the accuracy of multiple cutting-edge models into a single GW Bayesian analysis, while also reducing the computational cost (see Methods). This technique accounts for modelling errors by prioritising the most accurate GW model in each region of parameter space, thereby mitigating against biased results from using models that are unfaithful to general relativity. For GW signals likely observed by the LIGO\cite{TheLIGOScientific:2014jea}--Virgo\cite{acernese2014advanced}--KAGRA\cite{KAGRA:2020tym} GW detectors, we demonstrate that current techniques are more likely to inflate uncertainties and have the potential to produce biased parameter estimates. On the other hand, we show that the method presented here either outperforms current techniques, or in the worst case, gives comparable results.

\section{Gravitational-wave Bayesian inference}\label{sec:PE}


\begin{figure*}[t!]
  \centering
  \includegraphics[width=0.48\textwidth]{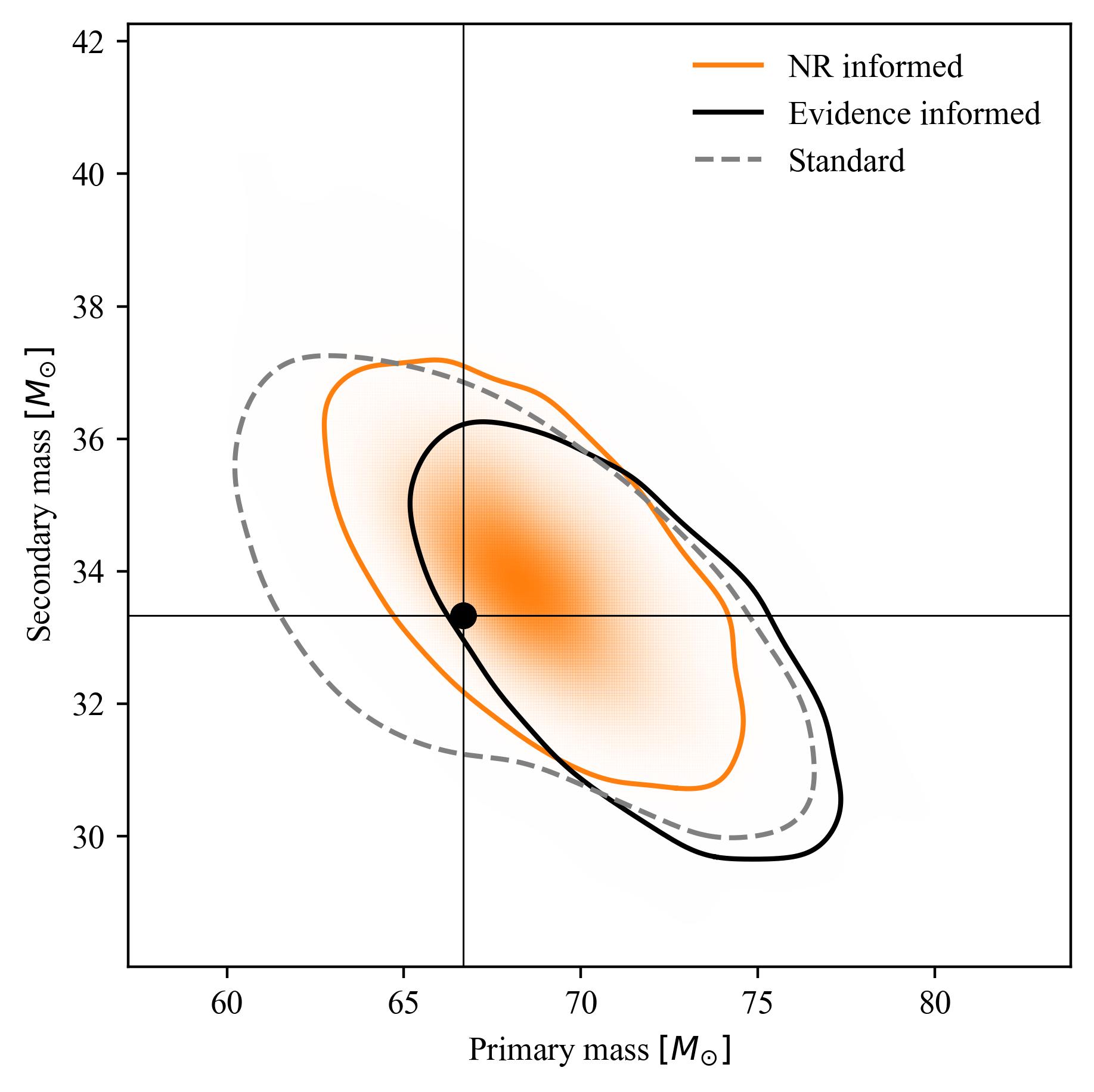}
  \includegraphics[width=0.48\textwidth]{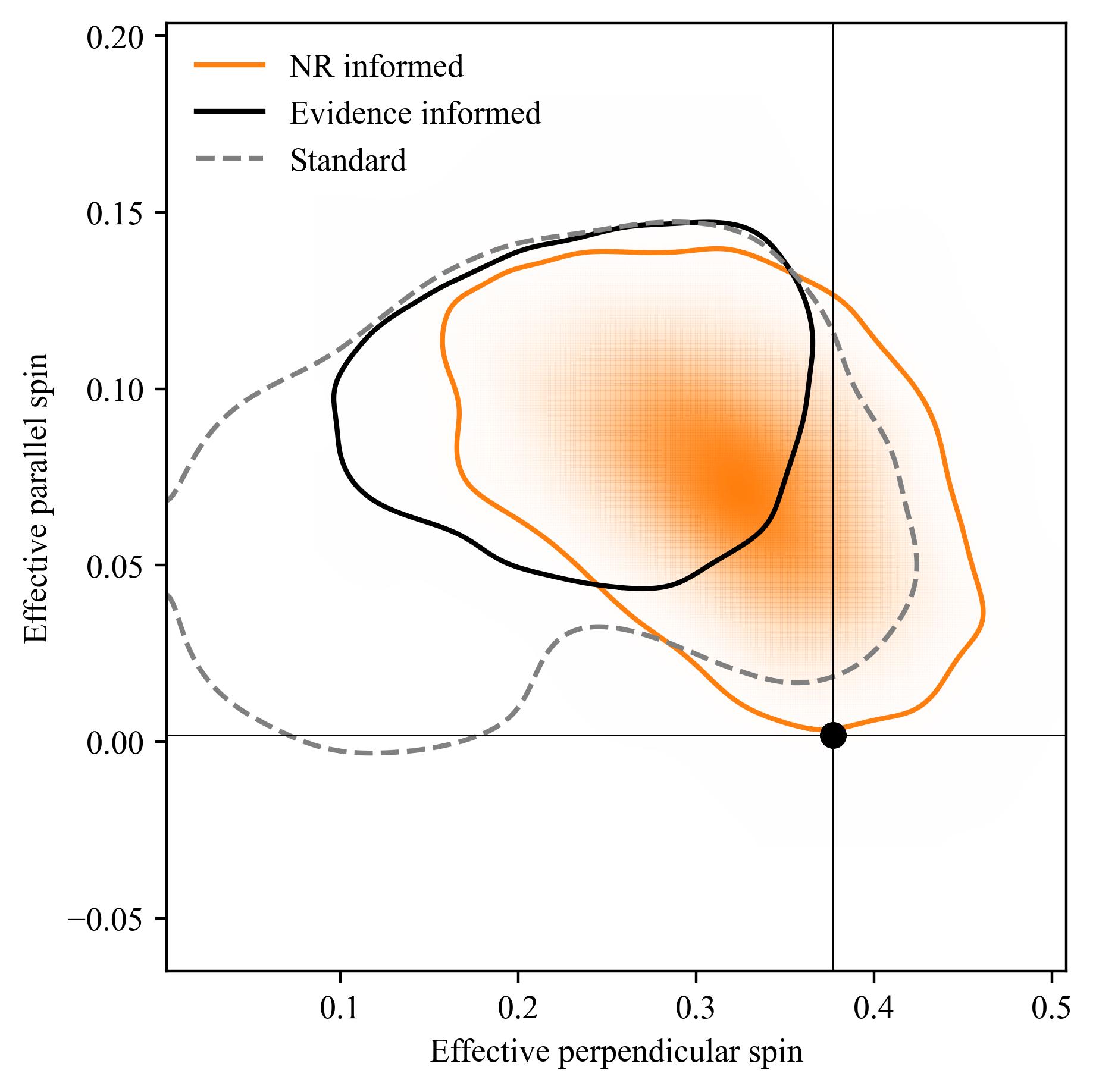}
  \caption{\textbf{Two-dimensional posterior probabilities obtained in our analysis of the {\texttt{SXS:BBH:0926}} numerical relativity simulation}. The left panel shows the measurement of the primary and secondary mass of the binary, and the right panel shows the inferred effective parallel and perpendicular spin components (as defined in Methods, see equations~(\ref{eq:chiperp}, \ref{eq:chi_par})). An effective perpendicular spin of 0 means that the spin vector lies perpendicular to the plane of the binary. The contours represent 90\% credible intervals and the black cross hairs indicate the true value.}
  \label{fig:pe_comparison}
\end{figure*}

We first apply our approach to analyse an example of a theoretical GW signal expected from general relativity, specifically, the {\texttt{SXS:BBH:0926}}\cite{Blackman:2017pcm,Boyle:2019kee} numerical relativity simulation produced by the Simulating eXtreme Spacetimes (SXS) collaboration (\href{https://www.black-holes.org}{https://www.black-holes.org}). We assume a total mass of $100\, M_{\odot}$ and we inject this signal into zero noise at a signal-to-noise ratio of 40. The {\texttt{SXS:BBH:0926}} simulation has mass ratio $1:2$ and large dimensionless spin magnitudes perpendicular to the orbital angular momentum (within the orbital plane of the binary) for both black holes of $\approx 0.8$ out of a maximum possible value of $1$.
For this system, the general relativistic phenomenon of spin-induced orbital precession\cite{Apostolatos:1994mx} is significant, and contributes a signal-to-noise ratio\cite{Fairhurst:2019vut} $\sim 9$ to the total power of the signal. This simulation was chosen since the majority of GW models obtain biased results, and disagree on the inferred binary parameters\cite{MacUilliam:2024oif,Akcay:2025rve}
Such a system with significant spin-induced orbital precession has been predicted to be observed once in every 50 GW observations made by the LIGO, Virgo and KAGRA GW observatories based on current black hole population estimates\cite{Hoy:2024qpy}.

We use three of the most accurate and cutting edge models currently available for describing the theoretical GW signals produced by colliding black holes:
{\modelname{IMRPhenomXPHM}}\cite{Pratten:2020ceb} (with the updated precession formulation\cite{Colleoni:2024knd}), 
{\modelname{IMRPhenomTPHM}}\cite{Estelles:2021gvs} and {\modelname{SEOBNRv5PHM}}\cite{Ramos-Buades:2023ehm}.~All 
models include the general relativistic phenomenon of spin-induced orbital precession\cite{Apostolatos:1994mx} and higher order multipole moments\cite{Goldberg:1966uu}. We analyse 8 seconds of data, and only consider frequencies between [20, 2048]\,Hz. We generate the numerical relativity simulation from $\approx 10\, \mathrm{Hz}$ to ensure that most higher multipole content is generated prior to our analysis window. Our analysis is restricted to a two-detector network of LIGO-Hanford and LIGO-Livingston\cite{TheLIGOScientific:2014jea}, and we assume a theoretical power spectral density for Advanced LIGO's design sensitivity\cite{dcc:T2200043}. We use the most agnostic priors available for all parameters, identical to those used in all detections made by the LIGO--Virgo--KAGRA collaboration\cite{LIGOScientific:2021djp}: flat in the component masses, spin magnitudes and cosine of the spin tilt angles. We perform Bayesian inference with the {\texttt{Dynesty}} Nested sampling software\cite{Speagle:2020} via {\texttt{Bilby}}\cite{Ashton:2018jfp}, as has been done in all LIGO--Virgo--KAGRA analyses since the third GW catalog\cite{LIGOScientific:2021djp}.

In Figure~\ref{fig:pe_comparison} we compare the results obtained with our method to two widely adopted techniques. The contours labelled \textit{NR informed} utilise the method presented here, \textit{Evidence informed} combines separate inference analyses obtained with different GW models according to their relative Bayesian evidence\cite{Ashton:2019leq}, and \textit{Standard} combines the results of separate inference analysis with equal weight. Standard is the currently adopted method by the LIGO--Virgo--KAGRA collaboration\cite{LIGOScientific:2021djp} as it is likely the most agnostic.
When considering the inferred primary and secondary masses of the binary, we see that all three techniques capture the true value within the two-dimensional marginalized 90\% credible interval. Both the NR Informed and Standard methods more accurately infer the true values of the binary, with the injected values lying within the 50\% credible interval. Given that the Standard method equally combines analyses from the individual GW models, the uncertainty is inflated in comparison to the method presented here and to the Evidence informed result.

We now turn our attention to the inferred spin on the binary. Since the individual spin components are difficult to measure for binary black holes at present-day detector sensitivities\cite{Purrer:2015nkh}, we consider the measurement of effective spin parameters that describe the dominant spin effects of the observed GW signal\cite{Ajith:2009bn,Ajith:2011ec}. In Figure~\ref{fig:pe_comparison} we show the measurement of the effective spin parallel and perpendicular to the orbital angular momentum, as defined below in Methods. We see significant differences between the obtained posterior distributions: the NR Informed approach introduced in this work is the least biased as it encompasses the true value within the two-dimensional marginalized 90\% credible interval. Although the Evidence informed result has been described as the optimal method in previous work\cite{Ashton:2019leq}, for this simulated signal it produces an inaccurate result. This is because {\modelname{IMRPhenomTPHM}} has the largest Bayesian evidence despite not being the most accurate model; it has been shown previously that less accurate models can give large Bayesian evidences due to mismodelling\cite{Hoy:2022tst}. Our analysis, on the other hand, predominantly uses {\modelname{SEOBNRv5PHM}}; specifically 90\% of the time, while {\modelname{IMRPhenomTPHM}} is used 8\% of the time, and {\modelname{IMRPhenomXPHM}} 2\% of the time. This demonstrates one of the limitations of our method: although we preferentially use the most accurate model in each region of parameter space, there is no guarantee that this model is accurate enough to avoid biases in the inferred parameter estimates\cite{Baird:2012cu,Toubiana:2024car}. However, we highlight that it is the most accurate method of those currently used, and can evolve to include more accurate models when they are developed.


\begin{figure*}[t!]
  \centering
  \includegraphics[width=0.98\textwidth]{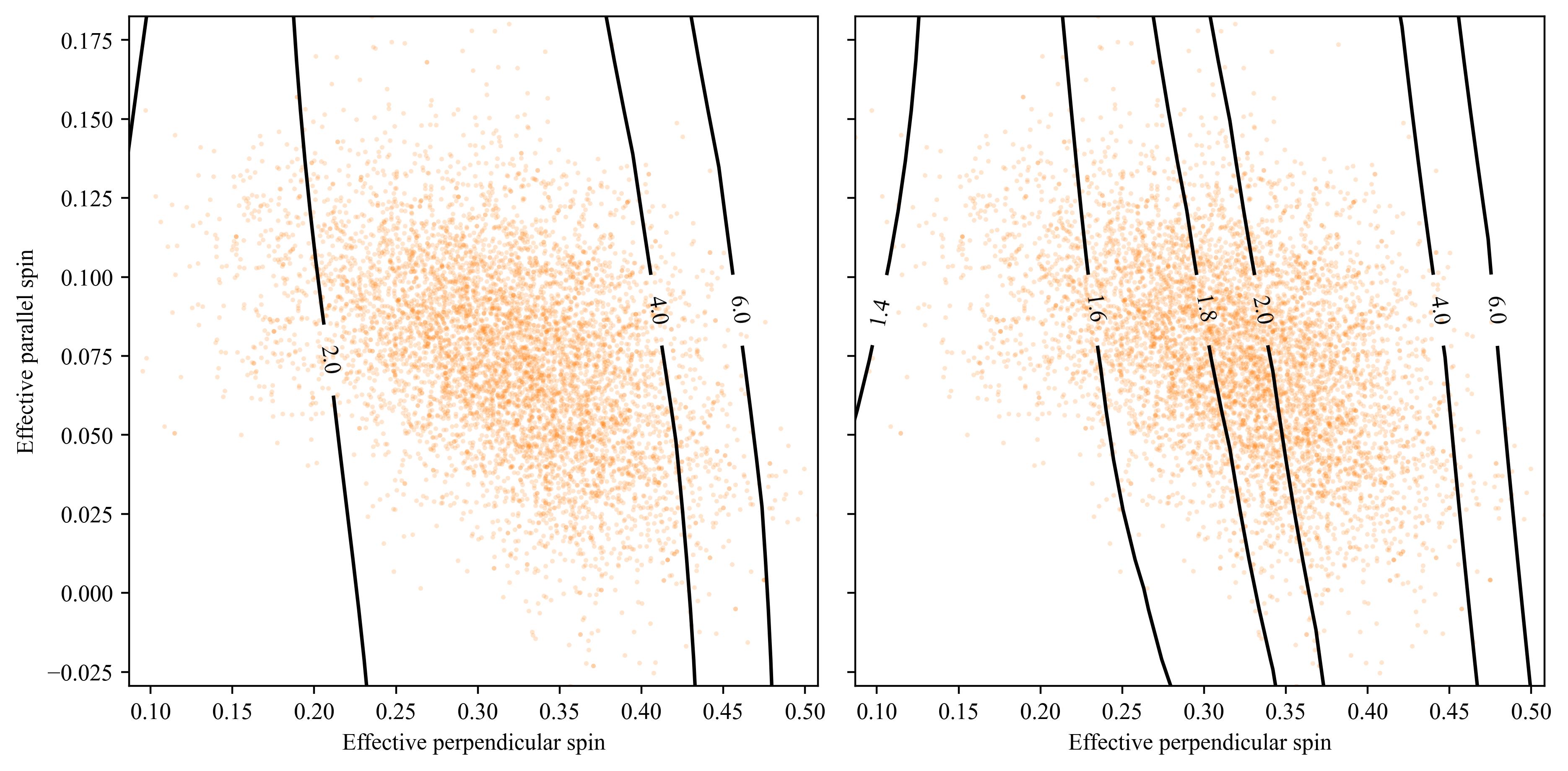}
  \caption{\textbf{Ratio of mismatches to numerical relativity simulations}. Contour plot showing the ratio of mismatches to numerical relativity simulations for different effective parallel and perpendicular spin components when averaging over different mass configurations. The left panel compares \modelname{IMRPhenomXPHM} and \modelname{SEOBNRv5PHM} and the right panel compares \modelname{IMRPhenomTPHM} and \modelname{SEOBNRv5PHM}. In orange we show samples obtained from our analysis of the {\texttt{SXS:BBH:0926}} numerical relativity simulation. In both cases, a ratio of mismatches greater than unity implies that {\modelname{SEOBNRv5PHM}} is more faithful to general relativity.}
  \label{fig:match_comparison}
\end{figure*}

In Figure~\ref{fig:match_comparison} we present the ratio of mismatches obtained with the different GW models used in this work. We see that {\modelname{SEOBNRv5PHM}} has the smallest mismatch in the region of parameter space containing the simulation parameters, and is therefore the most faithful to General Relativity results in this region. Specficially, it yields mismatches $\sim 3\times$ and $\sim 1.8\times$ smaller than {\modelname{IMRPhenomXPHM}} and {\modelname{IMRPhenomTPHM}}, respectively.

Since the NR informed approach chooses the GW model based on its accuracy to numerical relativity in each region of parameter space, rather than combining finalized results from each GW model individually, we also see a significant decrease in computational cost. Our analysis uses $30\%$ less computational resources than the Standard and Evidence informed analyses during sampling. The analysis completed in 230 CPU days, compared with 35 CPU days, 118 CPU days and 181 CPU days for the individual {\modelname{IMRPhenomXPHM}}, {\modelname{IMRPhenomTPHM}} and {\modelname{SEOBNRv5PHM}} analyses, respectively. In the worst case scenario, we expect our method to use the same computational resources as the Standard and Evidence informed analyses.

Our technique is free to use any combination of GW models. If {\modelname{SEOBNRv5PHM}} were removed from this analysis, we find consistent results between our method and the Evidence informed result, with overlapping two-dimensional marginalized 90\% confidence intervals. The reason is because {\modelname{IMRPhenomTPHM}} now has the largest Bayesian evidence and is the more accurate of the two remaining GW models considered in the region of parameter space.

A single analysis with the model that is, on average, the most accurate in the parameter space of interest can be performed\cite{Hannam:2021pit}. However, the issue with this technique is that the mismatch varies considerably across different regions of the parameter space, particularly for the spins which are often not well measured. For instance, when averaging across the parameter space consistent with {\texttt{SXS:BBH:0926}}, {\modelname{SEOBNRv5PHM}} is the most accurate model. However, for effective parallel spins $>0$ and perpendicular spins $< 0.05$, we find that {\modelname{IMRPhenomTPHM}} is more accurate than {\modelname{SEOBNRv5PHM}}, and {\modelname{IMRPhenomXPHM}} is of comparable accuracy to {\modelname{SEOBNRv5PHM}}. By simply averaging the mismatch across the parameter space, we neglect this information, resulting in the use of a less accurate model in certain regions of the parameter space.
On the other hand, the method presented in this work to incorporate the accuracy of multiple models into a single GW Bayesian analysis fully utilises this information.

Numerical relativity surrogate techniques provide accurate models for describing GWs produced from colliding black holes\cite{Varma:2018mmi,Varma:2019csw}. We do not sample over surrogate models in this work since they are used as a proxy for numerical relativity simulations to assess model accuracy (see Methods). We quantify the efficacy of our approach by comparing results to those obtained with surrogate models. For the same numerical relativity simulation we find that {\modelname{NRSur7dq4}} -- the leading generic-spin numerical relativity surrogate model\cite{Varma:2019csw} -- more accurately captures the true parameters of the binary as expected (see Supplementary Figure 1). Our NR informed approach offers the most statistically similar one-dimensional posterior probability distributions to the surrogate posteriors out of the methods considered in this work.

Contrary to standing belief,
{\modelname{NRSur7dq4}} \emph{is not guaranteed to be the most accurate model, even
within its calibration region}. For instance, when comparing against numerical relativity 
simulations that were not used to validate the {\modelname{NRSur7dq4}}, we find that
{\modelname{SEOBNRv5PHM}}. {\modelname{IMRPhenomTPHM}} and
{\modelname{IMRPhenomXPHM}} can more faithfully
describe numerical relativity than {\modelname{NRSur7dq4}}. Specifically, based on 
mismatches against the {\texttt{CF\_52}} simulation~\cite{Hamilton:2023qkv} -- a single spin,
mass ratio $1:4$ simulation with primary dimensionless spin magnitude $0.6$ at total masses
$75\, M_{\odot}$, $80\, M_{\odot}$ and $85\, M_{\odot}$ -- we estimate that according to our
NR informed approach, {\modelname{SEOBNRv5PHM}} would be $\approx 140$ times
more likely to be used than {\modelname{NRSur7dq4}} in this region of parameter space due
to its improved
accuracy. While corner cases such as this exist, {\modelname{NRSur7dq4}} is still suitable as a
proxy for numerical relativity in this work since we are performing Bayesian inference on 
numerical relativity simulations where the surrogate is the most accurate model~\cite{Varma:2019csw}.
We emphasise that once more numerical relativity simulations become available,
the surrogate will no longer be needed as a proxy to assess model accuracy, and we will
be able to incorporate the failthfulness of all models, including the surrogate, within
our Bayesian framework.

Although not presented in this section (see Supplementary Figures 2 and 3), we also analysed the {\texttt{SXS:BBH:0143}}\cite{Blackman:2017pcm,Boyle:2019kee} and {\texttt{SXS:BBH:1156}}\cite{Blackman:2017pcm,Boyle:2019kee} numerical relativity simulations produced by the SXS collaboration. {\texttt{SXS:BBH:0143}} was chosen since it resides in a region of the parameter space where we expect our method to give comparable results to the Standard and Evidence informed analyses. {\texttt{SXS:BBH:1156}} was chosen since it has largely asymmetric mass components and lies in the extrapolation regime of our technique (see Methods for details).
For the case of {\texttt{SXS:BBH:0143}}, we find largely overlapping posteriors between all three methods, with most of the one-dimensional marginalized 90\% confidence intervals containing the true value. Our analysis of this case uses {\modelname{SEOBNRv5PHM}} 80\% of the time, {\modelname{IMRPhenomTPHM}} 15\% of the time and {\modelname{IMRPhenomXPHM}} 5\% of the time. This represents the worst case scenario: by construction our method should at worst give the same results as other methods.
For the case of {\texttt{SXS:BBH:1156}}, we find that our method outperforms the Standard and Evidence informed analyses despite partly being in the extrapolation regime of our technique: we more accurately capture the true parameters of the binary. Similar to {\texttt{SXS:BBH:0926}}, the Evidence informed analysis preferred {\modelname{IMRPhenomTPHM}} owing to the larger Bayesian evidence, while our analysis preferred {\modelname{SEOBNRv5PHM}} since it is the more accurate model in this region of the parameter space. Our analysis used {\modelname{SEOBNRv5PHM}} 78\%, {\modelname{IMRPhenomTPHM}} 9\% and {\modelname{IMRPhenomXPHM}} 13\% of the time.


\begin{figure}[t!]
  \centering
  \includegraphics[width=0.48\textwidth]{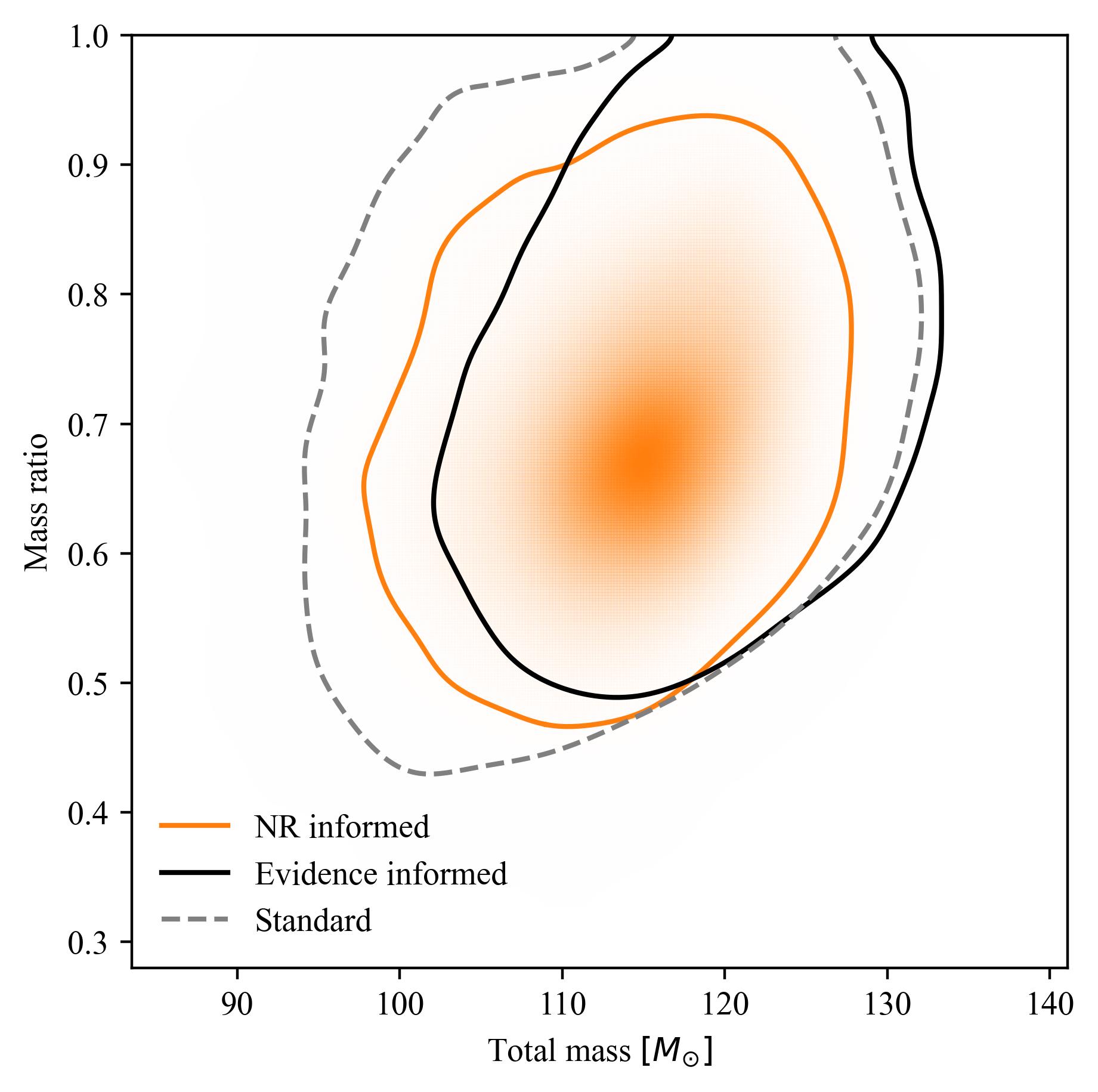}
  \caption{\textbf{Two-dimensional posterior probabilities obtained in our analysis of GW191109\_010717}. We show the measurement of the total mass of the binary, as well as mass ratio, defined as the secondary mass divided by the primary mass. A mass ratio of 1 implies equal binary component masses, and the mass ratio is always less than or equal to 1. The contours represent 90\% credible intervals.}
  \label{fig:GW191109_comparison}
\end{figure}

Finally, we apply our technique to a real gravitational-wave signal: GW191109\_010717 was observed on the 9 November 2019~\cite{LIGOScientific:2021djp}, and has sparked interest within the community since its source likely has large component masses that lie within the upper mass gap -- theories suggest that the maximum black hole mass from stellar collapse is $\sim 65\, M_{\odot}$~\cite{Woosley:2002zz}. As shown in Figure~\ref{fig:GW191109_comparison}, by incorporating model accuracy in GW Bayesian inference for the first time, we more tightly constrain the total mass of GW191109\_010717 to $100 < M < 124\, M_{\odot}$, and demonstrate that the source of GW191109\_010717 has conclusively unequal component masses (assuming the binary black hole hypothesis). Our re-analysis shows that when using consistent priors and sampler settings as the LIGO--Virgo--KAGRA collaboration, the primary component mass of GW191109\_010717 has a $69\%$ probability that it lies within the upper mass gap, consistent with previous work where GW191109\_010717 was re-analysed with {\modelname{NRSur7dq4}}~\cite{Islam:2023zzj}. This is compared to 51\% probability from the LIGO--Virgo--KAGRA analysis~\cite{LIGOScientific:2021djp}, thereby significantly increasing the probability that GW191109\_010717 was produced from a hierarchical formation mechanism where the primary component mass was formed from a previous black hole merger. Other one-dimensional posterior probability distributions remain comparable among the different methods considered in this work.

\section{Conclusions}
\label{sec:conclusion}

In this work we present a method to incorporate model uncertainty into
gravitational-wave Bayesian inference for the first time. We apply this method to theoretical GW signals expected from general relativity and show that it (i) marginalizes over model uncertainty by prioritising the most accurate model in each region of parameter space, and (ii) outperforms widely used techniques that use Bayesian model averaging. The method presented in this work is independent of the models chosen and can, in principle, be used with any combination.  Although the approach preferentially uses the most accurate model in each region of parameter space, there is no guarantee that this model is accurate enough to avoid biases in the parameter estimates. However, GW models are continuously being developed, and will likely improve in accuracy across the parameter space. Once available, these more accurate models can be incorporated into this method. Similarly, with more numerical relativity simulations produced, the accuracy of this method increases, and more models can be included. The method presented here is applicable to ground-based GW parameter estimation analyses and we highly encourage its use in the future.

\section{Methods}
\label{sec:method}

\subsection{Estimating waveform accuracy:}
\label{sec:matches}

As discussed in Section~\ref{sec:introduction}, the accuracy of a theoretical
GW model is often assessed by comparing the signals produced by the model against
numerical relativity simulations. We introduce a noise-weighted inner product
between the model representation of a signal and the signal itself\cite{Owen:1995tm},
\begin{equation}
  \langle h_\text{m}|h_\text{s} \rangle =
  4 \Re\int_{f_\text{min}}^{f_\text{max}}
  \frac{\tilde{h}_\text{m}^*\,\tilde{h}_\text{s}}{S_n(f)}\, \mathrm{d}f,
\end{equation}
where a tilde denotes a Fourier transform,
\({}^*\) denotes complex conjugation, and \(S_n(f)\) is the
noise power spectral density, which in this work is Advanced LIGO's design sensitivity\cite{dcc:T2200043}. The mismatch\cite{Owen:1995tm} between two signals is computed by
optimising the normalised inner product over a set of (intrinsic or extrinsic) model parameters \(\boldsymbol{\lambda}_\text{m}\),
\begin{equation}
  \label{eq:mismatch}
  \mismatch =
  1 - \max_{\boldsymbol{\lambda}_\text{m}} \frac{\langle h_\text{m}|h_\text{s}\rangle}{\sqrt{\langle h_\text{m}|h_\text{m}\rangle\langle h_\text{s}|h_\text{s}\rangle}}.
\end{equation}

The intrinsic parameter space for a generic quasi-circular compact binary system is comprised of
two masses, $m_{1,2}$, and two spin vectors, $\mathbf{S}_{1,2}$, adding up to eight degrees of freedom.
Additionally, a quasi-circular binary comes with seven more extrinsic parameters:
the right ascension, declination and the luminosity distance $\{\alpha,\delta,d_L\}$, respectively,
to the binary's center of mass; the inclination of the orbit and its relative polarisation
$\{\iota, \psi\}$; and overall constant time and phase shifts $\{t_\text{c},\varphi_\text{c}\}$ of the GW.

For binaries where the spins of the compact bodies are aligned with the system's orbital angular momentum, several of the
binary parameters become constant in time, and the intrinsic and extrinsic parameters decouple, thus reducing the dimensionality of
the model space to four degrees of freedom: \(\{m_1, m_2, \mathbf{S}_{1z}, \mathbf{S}_{2z}\}\), where the individual components of the spin vectors are specified at any fixed frequency. To compute the matches for the aligned-spin configurations that follow, we hold all extrinsic parameters fixed and optimise the match over the set of model parameters \(\boldsymbol{\lambda}_\text{m}=\{t_\text{c},\varphi_\text{c}\}\). We maximise over \(t_\text{c}\) via an inverse fast Fourier transform and over \(\phi_\text{c}\) using the Nelder-Mead optimization algorithm found in \textsc{SciPy}'s \texttt{minimize} function\cite{SciPy-NMeth:2020aaa}.

For binary systems in which the spins contain non-zero components orthogonal to the orbital angular momentum, the intrinsic and extrinsic parameters couple and evolve in time. Our aim is to isolate the intrinsic parameter space,
accordingly the mismatch to which we intend to fit should somehow be independent of the extrinsic parameters.
For this purpose, we first map $\{\alpha,\delta,\psi\}$ into a single parameter known as the effective polarisability\cite{Harry:2016ijz} $\kappa$.
We then prepare an evenly-spaced signal grid over the $\{\kappa, \varphi_\text{c},\iota\}_\text{s}\in [0, \pi/2]\otimes [0, 2\pi)\otimes [0,\pi]$ space with $7\times 6\times 7=294 $ elements.
At each point in this signal grid, we compute
the sky-optimised mismatch\cite{Harry:2016ijz, Gamba:2021ydi, MacUilliam:2024oif}
between the signal and the model template from equation~\eqref{eq:mismatch}, where the parameter set we optimise over is \(\boldsymbol{\lambda}_\text{m}=\{t_\text{c}, \varphi_\text{c}, \kappa,\varphi_\text{spin}\}\). Here
$\varphi_\text{spin}$ represents the freedom to rotate the in-plane spin (azimuthal) angles $\phi_1, \phi_2$ of $\mathbf{S}_1,\mathbf{S}_2 $ by a constant amount.
$\kappa$ is optimised over analytically and optimisations over $\varphi_\text{c}$ and $\varphi_\text{spin}$ are performed numerically using
dual annealing algorithms\cite{Pratten:2020ceb, Gamba:2021ydi, Thompson:2023ase}.
Note that there is no universally agreed upon grid for
$\{\kappa, \varphi_\text{c}\}$\cite{Khan:2018fmp, Gamba:2021ydi, Colleoni:2024knd},
nor for $\iota$\cite{Hamilton:2021pkf, Thompson:2023ase}.
Our specific choice for the $\{\kappa, \varphi_\text{c}\}$ grid is based on recent work\cite{MacUilliam:2024oif}.
Our $\iota$ grid spacing is also consistent with the literature\cite{Hamilton:2021pkf},
but extended to $\pi$ because the ``up-down'' symmetry of the GW multipoles with respect to the
orbital plane is broken due to precession\cite{Bruegmann:2007bri, Keppel:2009tc, Kalaghatgi:2020gsq, Ramos-Buades:2020noq, Ghosh:2023mhc, Kolitsidou:2024vub}.

With these optimisations, we arrive at the maximum possible match between the template and the signal
at a given point $\{\kappa, \varphi_c,\iota\}_\text{s}$ in the signal grid.
We repeat this procedure at every point of the 294-element grid then compute the mean
of this set as our final result for the mismatch
\be
\mismatch_\text{av}:=\f{1}{294}\sum_{\text{s}=1}^{294}\mismatch(\kappa_\text{s},\varphi_\text{c,s},\iota_\text{s})\label{eq:M_opt_av}.
\ee
This is done to marginalise over any dependence of the mismatch on the sky position and inclination,
thus obtaining values which depend exclusively on the intrinsic parameters of the source.
We additionally retain the standard deviation $\sigma$ of the 294-mismatch set
and use this as our error bar when needed.
Note that our mean match, $1-\mismatch_\text{av}$, is a discretely averaged version of
the sky-and-polarisation averaged faithfulness given by equation~(35) of Ramos-Buades~\emph{et al.}\cite{Ramos-Buades:2023ehm}.
For the remainder of this article, we drop the subscript ``av'' from $\mismatch$.

\subsection{Multi-model Bayesian inference:}

The parameters of a binary are inferred from a gravitational-wave signal through Bayesian inference. Here, the model dependent posterior distribution for parameters $\boldsymbol{\lambda} = \{\lambda_{1}, \lambda_{2}, ..., \lambda_{j}\}$ is obtained through Bayes' theorem,

\begin{equation}
  p(\boldsymbol{\lambda} \vert d, \model_{i}) =  \frac{\Pi(\boldsymbol{\lambda} \vert \model_{i})\, \mathcal{L}(d \vert \boldsymbol{\lambda}, \model_{i})}{\mathcal{Z}},
\end{equation}
where  $\Pi(\boldsymbol{\lambda} \vert \model_{i})$ is the probability of the parameters $\boldsymbol{\lambda}$ given the model $\model_{i}$, otherwise known as the prior; $\mathcal{L}(d \vert \boldsymbol{\lambda}, \model_{i})$ is the probability of observing the data given the parameters $\boldsymbol{\lambda}$ and model $\model_{i}$, otherwise known as the likelihood; and $\mathcal{Z}$ is the probability of observing the data given the model $\mathcal{Z} = \int{\Pi(\boldsymbol{\lambda} \vert \model_{i})\, \mathcal{L}(d \vert \boldsymbol{\lambda}, \model_{i})\, d\boldsymbol{\lambda}}$, otherwise known as the evidence. It is often not possible to trivially evaluate the model dependent posterior distribution; the challenge is evaluating the evidence since it involves computing the likelihood times prior for all points in the parameter space. Thankfully, nested sampling was developed to estimate the evidence through stochastic sampling and return the model dependent posterior distribution as a by-product\cite{Skilling:2006}. Here, a set of \textit{live points} are randomly drawn from the prior, and the point with the lowest likelihood is stored and replaced with another point randomly drawn from the
\emph{likelihood-constrained prior}; the new point is randomly drawn from the prior provided that the likelihood is larger than the point that it is replacing.
This iterative process continues until the highest likelihood region(s) is identified.

When there is an ensemble of models, Bayesian model averaging can be used to marginalize over the model uncertainty,

\begin{align} \label{eq:multi_waveform_bayes_theorem}
  p(\boldsymbol{\lambda} \vert d) & = \sum_{i=1}^{N} p(\boldsymbol{\lambda} \vert d, \model_{i}) \,p(\model_{i} \vert d) \nonumber                                                                         \\
                                  & = \sum_{i=1}^{N}{\left[\frac{\mathcal{Z}_{i}\, \Pi(\model_{i})\,p(\boldsymbol{\lambda} \vert d, \model_{i})}{\sum_{j=1}^{N} \mathcal{Z}_{j}\,\Pi(\model_{j})}\right]},
\end{align}
where $p(\model_{i} \vert d) $ is the probability of the model $\model_{i}$ given the data, $\Pi(\model_{i})$ is the discrete prior probability for the choice of model, and $N$ is the number of models in the ensemble. For the case of uniform priors for the model, i.e., $\Pi(\model_{i}) = 1 / N$, Bayesian model averaging simply averages the model-dependent posterior distributions, weighted by the evidence.

An alternative solution to marginalizing over model uncertainty is to simultaneously infer the model and model properties in a single \textit{joint} analysis\cite{Hoy:2022tst}. Here, the parameter set $\boldsymbol{\lambda}$ is expanded to include the model $m$: $\tilde{\boldsymbol{\lambda}} = \{\lambda_{1}, \lambda_{2}, ..., \lambda_{j}, m\}$, and a discrete set of models can be sampled over during standard Bayesian inference analyses: for each step in, \textit{e.g.}, a nested sampling algorithm, a ($j$+1)-dimensional vector of model parameters is drawn from the prior,
including an integer for the model, $m$. The integer $m$ is mapped to a gravitational-wave model, and the likelihood is evaluated by passing the remaining model parameters, and the selected model, to the standard gravitational-wave likelihood\cite{Veitch:2014wba}. It was demonstrated that this joint analysis will be at most $N \times$ faster to compute compared to performing Bayesian model averaging\cite{Hoy:2022tst}.

For the case of gravitational-wave astronomy, defining a discrete prior probability for the model is challenging since the accuracy of each model varies across the parameter space $\boldsymbol{\lambda}$\cite{MacUilliam:2024oif}. This makes it difficult to perform Bayesian model averaging; a uniform prior probability is often assumed for the choice of model\cite{Ashton:2019leq,Jan:2020bdz} or in some cases, the model accuracy is averaged over the parameter space of interest\cite{Hoy:2022tst}. However, a parameter-space dependent prior for the choice of model may solve this problem\cite{Hoy:2022tst}. For instance, a $j$-dimensional vector of model parameters can be drawn from the prior and $\Pi(\model_{i} \vert \boldsymbol{\lambda})$ can be evaluated for all models, i.e., the prior probability of the model given the parameter set $\boldsymbol{\lambda}$. The most probable model can then be determined, and the gravitational-wave likelihood subsequently evaluated. Although other priors have been suggested\cite{Hoy:2022tst,Jan:2021aaa}, we use the following model prior conditional on the parameters $\boldsymbol{\lambda}$,

\begin{equation} \label{eq:model_prior}
  \Pi(\model_{i} \vert \boldsymbol{\lambda}) = \frac{\mismatch_{i}(\boldsymbol{\lambda})^{-4}}{\sum_{j} \mismatch_{j}(\boldsymbol{\lambda})^{-4}}
\end{equation}
where $\mismatch(\boldsymbol{\lambda})$ is the mismatch between the model $\model_{i}$ and a numerical relativity simulation with parameters $\boldsymbol{\lambda}$. Equation~\eqref{eq:model_prior} implies that the most accurate GW model will more likely be used to evaluate the likelihood in each region of parameter space. 

While several mismatch-dependent priors were tested, equation~\eqref{eq:model_prior} was chosen for this work since it accentuates small differences in the mismatch between models, and was found to perform optimally. However, since the mismatch is a function of the power spectral density, it will subtly change when the profile of the power spectral density is varied, for example due to an improvement in the sensitivity of GW detectors as a result of commissioning periods or due to small variations on a day-to-day basis from noise artefacts. As a result, it is possible that the relative model probabilities in equation~\eqref{eq:model_prior} will vary for different power spectral density realisations. We note that this is a common problem in GW astronomy with, \textit{e.g.}, search pipelines similarly using a single representative power spectral density when constructing template banks\cite{DalCanton:2017ala}. We leave a more detailed analysis investigating the choice and stability of this distribution to future work.

\subsection{Constructing a match interpolant:}
\label{sec:interpolant}

Mismatch computations are fast, taking $O(\text{ms})$ per evaluation,
for simplified models of the GW signal, such as those for aligned-spin configurations with only dominant quadrupolar emission.
With increased model complexity, the computation can take a significantly longer time to evaluate, and producing the mismatch $\mismatch(\boldsymbol{\lambda})$ will be a limiting cost in a Bayesian analysis since the likelihood is evaluated $O(10^{8})$ times during a typical nested sampling analysis. For this reason, we construct an interpolant for the mismatch across the parameter space, $\mismatch(\boldsymbol{\lambda})$,
based on a discrete set of $K$ mismatches for each of the GW models used in this analysis. 

Owing to computational limitations, we do not have numerical relativity simulations for all possible regions of the compact binary parameter space. For the aligned-spin interpolant construction we therefore evaluate mismatches using the numerical
relativity hybrid surrogate model \modelname{NRHybSur3dq8}\cite{Varma:2018mmi} as a
proxy for numerical relativity simulations. There is a long and productive history of GW signal modelling using a variety of approaches\cite{Bohe:2016gbl, Cotesta:2018fcv, Cotesta:2020qhw, Ossokine:2020kjp,Babak:2016tgq,Pan:2013rra, Husa:2015iqa, Khan:2015jqa, London:2017bcn, Hannam:2013oca, Nagar:2018zoe, Khan:2018fmp, Khan:2019kot, Pratten:2020fqn, Garcia-Quiros:2020qpx, Pratten:2020ceb, Estelles:2020osj, Estelles:2020twz, Estelles:2021gvs, Gamba:2021ydi, Thompson:2023ase, Nagar:2023zxh}, and we compare against the
  {\modelname{IMRPhenomXHM}}\cite{Garcia-Quiros:2020qpx} and
  {\modelname{IMRPhenomTHM}}\cite{Estelles:2020twz} waveform models,
two of the leading frequency and time-domain models available
for aligned-spin binaries, respectively.
We do not use the state of the art EOB models\cite{Ramos-Buades:2023ehm, Nagar:2023zxh}
for the aligned-spin proof-of-principle test because \modelname{IMRPhenom} models are one to two orders of magnitude faster to evaluate.
For the precessing model interpolants we use the models described in Section~\ref{sec:PE}: {\modelname{IMRPhenomXPHM}}\cite{Pratten:2020ceb} (with the updated precession formalisation\cite{Colleoni:2024knd}), {\modelname{IMRPhenomTPHM}}\cite{Estelles:2021gvs} and {\modelname{SEOBNRv5PHM}}\cite{Ramos-Buades:2023ehm}, and we compare the precessing models against the numerical relativity waveform surrogate model {\modelname{NRSur7dq4}}\cite{Varma:2018mmi,Varma:2019csw} as a proxy for full numerical relativity simulations when computing mismatches.

We next describe how we construct an interpolant for binaries with spins aligned with the orbital angular momentum in Section~\ref{sec:aligned_interpolant}. We further test this interpolant by comparing the posterior samples obtained
from a Bayesian inference analyses that is guided by an actual mismatch computation at every step vs. a Bayesian inference analyses guided by the interpolant. In Section~\ref{sec:prec_interpolant} we describe how we generalise this to build a generic spin interpolant. Due to computational cost, we use the Bayesian inference verification analysis in Section~\ref{sec:aligned_interpolant} to justify using an interpolant-guided analysis for systems with generic spins.

\subsection{Interpolant for aligned-spin waveform mismatches:}
\label{sec:aligned_interpolant}
We begin with a test of the method using aligned-spin gravitational wave models
containing higher signal multipoles.
To simplify the construction of the mismatch interpolant for this test application,
we reduce the dimensionality of the mismatch parameterisation by artificially
fixing several signal and model parameters. We choose to fix
the total mass of the binary to  \(M=90 M_\odot\) and
the inclination angle to \(\theta_\text{JN}=\pi/3\), where $\theta_\text{JN}$ spans the angle between
the line of sight to the binary and the total angular momentum vectors.
This choice leaves three remaining free parameters
in each model: the mass ratio \(q=m_2/m_1\leq1\)
and the component spins of
the primary and secondary masses aligned with the orbital angular momentum,
\(\chi_1\) and \(\chi_2\), respectively, defined from
$ \chi_i = \mathbf{S}_{iz}/m_i^2$ for $i=1,2$ with $-1\le \chi_i\le 1$.

The 3-dimensional mismatch interpolants are constructed from mismatches computed on
a uniform grid of 8 points in \(0.125\leq q\leq1\) and 17 points in each
\(-0.8\leq\chi_{1,2}\leq 0.8\), providing 2312 total mismatch points
for each model. The interpolants are produced as polynomial fits to \(\log_{10}\mismatch\) of the form
\begin{equation}
  \label{eq:aligned_spin_interpolant}
  \log_{10}\mismatch(q,\chi_1,\chi_2)=\sum_{\substack{0\leq a \leq 6 \\ 0\leq b,c \leq 8}} f_{abc} \,q^a\chi_1^b\, \chi_2^c,
\end{equation}
with the fit coefficients \(f_{abc}\) computed
using \textsc{Mathematica}'s \texttt{Fit} function and exported to \textsc{Python}
using \texttt{FortranForm}. These mismatch surfaces are well-behaved and we find that the simple polynomial fits described provide sufficiently small relative errors (arising from equations~\eqref{eq:rel_diff1} and~\eqref{eq:rel_diff2} described below) of \(10^{-4}\) and \(10^{-3}\), respectively, which suffices for this initial proof-of-principle test.

Next, we validate that our interpolant gives indistinguishable results to computing the mismatch directly in a Bayesian inference analysis. We perform two Bayesian inference analyses, both with the {\texttt{Dynesty}} Nested sampling software\cite{Speagle:2020} via {\texttt{Bilby}}\cite{Ashton:2018jfp}. We use the same priors and sampler settings as those typically used in LIGO--Virgo--KAGRA analyses. The only distinguishing
factor between these runs is that in one we use
equation~\eqref{eq:aligned_spin_interpolant} when computing
the conditional probabilities of equation~\eqref{eq:model_prior}, and in the other we directly compute the mismatch between the models and the surrogate at the sample point.

To compare posterior distributions we use the Jensen-Shannon Divergence\cite{Lin:1991zzm} since it is commonly used in gravitational-wave astronomy\cite{LIGOScientific:2018mvr,LIGOScientific:2020ibl}. The Jensen-Shannon Divergence ranges between $0\, \mathrm{bits}$, statistically identical distributions, and $1\, \mathrm{bits}$, statistically distinct distributions. A general rule of thumb is that a Jensen-Shannon Divergence $< 50\,\mathrm{mbits}$ implies that the distributions are in good agreement\cite{LIGOScientific:2018mvr}.

In Supplementary Table 1 we present Jensen-Shannon Divergences between marginalized posterior distributions obtained when a) calculating the mismatch exactly, and b) using the interpolant. We find that all Divergences are significantly less than $50\, \mathrm{mbits}$, implying that the distributions are close to statistically identical. We find that the Bayesian analysis that used the interpolant completed in $\sim 500$ CPU hours, $\sim 250\times$ faster than the Bayesian analysis that computed the mismatch exactly. Given the almost statistically identical posteriors and reduced computational cost, we use the interpolated mismatch for all subsequent analyses.

\subsection{Interpolant for precessing waveform mismatches:}
\label{sec:prec_interpolant}

When computing interpolants for the mismatches in equation~\eqref{eq:M_opt_av}, we choose to fit for the $\log_{10}$ of the sky-averaged, optimized waveform mismatch equation~\eqref{eq:M_opt_av}.
Accordingly, our error bars become
$\sigma_\text{log} :=\vert \log_{10}(\mismatch-\sigma) -
  \log_{10}(\mismatch+\sigma)\vert$.

Next, we generate a mismatch dataset to be used for fit construction (training).
We could simply select values for the intrinsic parameters $\{m_1,m_2,\mathbf{S}_1,\mathbf{S}_2\}$
and obtain $\mismatch$ via the procedure above, but we find that the brute force use of analytic functions of eight variables to fit to this data set is not the best approach.
Instead, we opt to first reduce the dimensionality of the parameter space and then employ functional fitting.
Already in Appendix~A of MacUiliam~\emph{et al.}\cite{MacUilliam:2024oif}
we had seen encouraging preliminary results of this approach.
We also note that generating just a single data point for this mismatch set is computationally expensive
because of the four-dimensional optimisation over $\boldsymbol{\lambda}_\text{m}=\{t_\text{c},\phi_\text{c}, \kappa,\varphi_\text{spin}\}$
that needs to be repeated for every element of the 294-term sum in equation~\eqref{eq:M_opt_av}.
For example, depending on mass ratio and total mass, the computation of the average mismatch equation~\eqref{eq:M_opt_av}
at a single point in the intrinsic parameter space takes approximately 2-3 CPU hours for \modelname{IMRPhenomXPHM}, 4.5-11 CPU hours for \modelname{IMRPhenomTPHM}, and 6 CPU hours for \modelname{SEOBNRv5PHM}.
Therefore, we must keep in mind computational economics when generating the data to construct the fits.

We start by mapping $m_{1,2}$ to the total mass $M$ and the symmetric mass ratio
$\eta$ via
\be
\Mtot = m_1+m_2,\quad \eta:= \f{m_1 m_2}{\Mtot^2} \label{eq:Mtot_eta}
\ee
with the former quoted in solar masses ($\Msun$) here and the latter being bounded $0< \eta\le 1/4$.
Alternatively, we could have worked with the chirp mass $M_c:=(m_1 m_2)^{3/5}(m_1+m_2)^{-1/5}$
instead of $\Mtot$, but we opt to work with the total mass as its impact on the mismatch, equations~(\ref{eq:mismatch}, \ref{eq:M_opt_av}) has been well documented\cite{Pratten:2020fqn, Estelles:2021gvs, Gamba:2021ydi, Ramos-Buades:2023ehm, MacUilliam:2024oif}.

The Cartesian components of each spin vector may be written in terms of spherical coordinates with respect
to some reference frame, usually taken to be the orbital angular momentum vector at a reference frequency\cite{Schmidt:2017btt}.
Thus, we may write $\mathbf{S}_{i}=\vert \mathbf{S}_{i}\vert(\sin\theta_i\cos\phi_i, \sin\theta_i\sin\phi_i ,\cos\theta_i)^T$
for $i=1,2$.

We reduce the dimensionality of this eight-dimensional intrinsic parameter space
by mapping the six-dimensional spin space to two effective spins that we here
label as $x$ and $y$, representing the effective spin projections perpendicular and parallel
to the reference orbital angular momentum vector of the binary, respectively.
Two logical candidates for $\{x,y\}$ already exist:
$\{\chi_\text{p},\chi_\text{eff}\}$.
The former is given by\cite{Hannam:2013oca, Schmidt:2014iyl}
\begin{equation}
  \chip=\max\left(\bar{S}_1\sin\theta_1,q\,\frac{4q+3}{4+3q}\,\bar{S}_2\sin\theta_2\right) \label{eq:chi_p}
\end{equation}
with the bounds $ 0\le \chip \le 1$, and we have introduced \(\bar{S}_{1,2}=|\mathbf{S}_{1,2}|/m_{1,2}^2\).
A non-zero value for this quantity is an indication of spin precession, with $\chip = 1$ corresponding
to a maximally precessing binary, \textit{i.e.}, all component spins of the binary constituents lie in the orbital plane and take their maximum magnitudes.

$\chieff$ is the parallel projection counterpart to $\chip$. It reads\cite{Damour:2001tu, Racine:2008qv, Hannam:2013oca, Purrer:2013ojf}
\be
\chi_\text{eff} = \f{1}{1+q}(\chi_1+q\chi_2) = \f{1}{1+q}(\bar{S}_1 \cos\theta_1+ q \bar{S}_2\cos\theta_2) \ .
\label{eq:chi_eff}
\ee
This is a conserved quantity up to 1.5 post-Newtonian (PN) order\cite{Racine:2008qv} and its magnitude changes very little
over the course of an inspiral, making it very useful for inferring spin information about a compact binary
system. It is clear from equation~\eqref{eq:chi_eff} that $-1\le \chieff \le 1$ given
the Kerr spin limit $|\chi_{1,2}| \le 1$.

Other perpendicular projections exist in the literature\cite{Thomas:2020uqj, Hamilton:2021pkf}, but the one which we empirically determined to be
the best for fitting is $\chiperp$ given by\cite{Akcay:2020qrj}
\begin{equation}
  \chiperp=\frac{\vert \mathbf{S}_{1,\perp}+\mathbf{S}_{2,\perp} \vert}{M^2},\label{eq:chiperp}
\end{equation}
where $\mathbf{S}_{i,\perp}=\bar{S}_i m_i^2(\sin\theta_i\cos\phi_i, \sin\theta_i\sin\phi_i ,0)^T$ for $i=1,2$.
We also experimented with a generalized version of $\chip$\cite{Gerosa:2020aiw},
but found this quantity to be not as well suited for fitting as $\chip$ or $\chiperp$.
Given that the mismatches will be maximized over the in-plane spin angle $\varphi_\text{spin}$,
we map $\phi_{1,2}$ to a single azimuthal spin angle $\Delta\phi=\phi_2-\phi_1$ by rotating our source frame axes such that $\phi_1=0$.

Finally, as an alternative to $\chieff$, we introduce
\be
\chi_\parallel := \frac{\vert \mathbf{S}_{1,\parallel}+\mathbf{S}_{2,\parallel} \vert}{M^2}=\f{1}{(1+q)^2}(\chi_1+q^2\chi_2) \ .
\label{eq:chi_par}
\ee

We thus have several choices for each perpendicular/parallel scalar:
$ x=\chip\text{ or }\chi_\perp  ,y=\chieff\text{ or }\chipar$,
yielding four possible pairings for the dimensional reduction of the spin space.
Our preliminary work based on gauging the faithfulness of the fits has, however,
compelled us to drop $\chip$ as it produced less faithful results,
partly due to the fact that it does not carry any information about the planar spin angle separation $\Delta\phi$.
Thus, we are left with two possible pairings for the reduced spin space:
$\{\chiperp,\chieff\}$ and $\{\chiperp, \chipar\}$.
Accordingly, we introduce the fit training-set labels $K_1=\{\chiperp, \chieff,\eta,\Mtot\}$ and
$K_2=\{\chiperp, \chipar,\eta,\Mtot\}$.

The spin parameters introduced above depend on $q$, accordingly our fitting variables $\{x,y,\eta\}$
do not form a linearly independent three-dimensional subspace.
Our motivation for choosing the particular fit variables above was ultimately empirical:
our initial fits, using projections of spins with no $q$ dependence, were less faithful to the data.
It seems that mass-ratio dependent spin projections retain more useful information
when the dimensionality of the parameter space is reduced.
Additionally, we find that $\{x,y,\eta\}$ are either not correlated or weakly
correlated for which we present correlation coefficients at the end of this section.

Next, we introduce a discrete parameter grid over the chosen four-dimensional
$\{\chiPerp, \chiPar,\eta,\Mtot\}$ space that we use for fitting.
We limit \(\eta\) to range from $\eta=0.16$ (corresponding to $q=1/4$) to $\eta=0.25$ ($q=1$) in four
even steps, resulting in five distinct values $\eta_j$, $j=1,\ldots 5$.
For the total mass, we employ $\Mtot=\{75,117.5,150\}\Msun$ as our grid points, chosen because
\begin{inparaenum}[(i)]
  \item \NRsur{} has been trained with data from binaries with only $q\ge 1/4$,
  \item \NRsur's time length limit\cite{Varma:2019csw} of $4300M$ imposes $\Mtot \gtrsim 75\Msun$ in order for the
  binary to enter the detector bandwidth at a GW frequency of $20\,$Hz,
  \item binaries with $\Mtot > 150\Msun$ mostly emit merger-ringdown signals in the detection band\cite{MacUilliam:2024oif},
  thus leaving hardly any imprint of precession in the reconstructed waveform from detector data,
  \item model mismatches tend to weakly depend on the total mass\cite{Pratten:2020fqn, Estelles:2021gvs, Gamba:2021ydi, Ramos-Buades:2023ehm, MacUilliam:2024oif}\!,
  thus sufficing three grid points in mass space for our current purposes given the computational burden
  of generating new data.
\end{inparaenum}

For better fit performance, the remaining two fit parameters, $\chiPerp$ and $ \chiPar$,
should also be placed on a regular grid. However, the quantities that we picked to cover this space,
namely the pairings $\{\chiperp,\chieff\}$ and $\{\chiperp, \chipar\}$ are not intrinsic parameters
of the binary system. In order to construct a regular grid in $\{\chiPerp, \chiPar\}$,
we therefore start from first a regular grid of roughly 50,000 elements in
$\{\bar{S}_1,\bar{S}_2,\theta_1,\theta_2,\Delta\phi\}$ space and use this to populate
the $\{\chiPerp, \chiPar\}$ space with values of $q$ already determined by the $\eta_j$ grid.
The resulting grid in, \textit{e.g.}, the $\chiperp$-$\chieff$ plane is scatter-plotted as the blue dots
in the left panel of Supplementary Figure 4, where we observe that
the parameter space seems to be bounded by a half \emph{prolate} ellipse drawn as the orange curve.
The horizontal/vertical axes of the ellipse are given by
\be
a=\text{max}(\chiPerp),\quad b=\text{max}(\chiPar)\label{eq:ellipse_axes}.
\ee
Guided by this observation, we construct a regular, \emph{elliptical} grid in $\{\chiPerp, \chiPar\}$
space as follows. First, we introduce the elliptical coordinates $A,\Phi$ with oblate/prolate-ness parameter $\mu>0$
\begin{subequations}
  \begin{align}
    x & = A \sinh \mu \cos\Phi, \label{eq:ellipse_x} \\
    y & = A \cosh\mu \sin\Phi \label{eq:ellipse_y}
  \end{align}
\end{subequations}
with $\Phi \in[0,2\pi]$ and the usual parametrization
\be
\f{x^2}{A^2\sinh^2\mu} + \f{y^2}{A^2\cosh^2\mu} = 1 \label{eq:ellipse_parametrization}.
\ee
For an ellipse of fixed size,
$A$ and $\mu$ are obtained from the relations $A \sinh\mu = a, A\cosh\mu=b$.
Note that in equations~(\ref{eq:ellipse_x}-\ref{eq:ellipse_parametrization}),
we swapped $\cosh\mu$ and $\sinh\mu$ because, as we show below,
our ellipses are prolate, i.e., $a<b$.

Here, we aim to create a grid based on ``concentric'' ellipses of the same aspect ratio starting with
the outermost one (orange curve in the left panel of Supplementary Figure 4).
With $A$ and $\mu$ fixed, we create an elliptical grid of our choosing via
\begin{subequations}
  \begin{align}
    x_{rs} & = \f{r}{N_r}\, A \sinh \mu \cos\left(\f{\pi}{N_s}s-\f{\pi}{2} \right), \label{eq:ellipse_grid_x} \\
    y_{rs} & =\f{r}{N_r}\, A \cosh \mu \sin\left(\f{\pi}{N_s}s-\f{\pi}{2} \right)\label{eq:ellipse_grid_y}
  \end{align}
\end{subequations}
with  $r=1,\ldots, N_r$ and $ s=0,\ldots N_s $.
We show such a grid for $\{x,y\}=\{\chiperp, \chieff\}$ in the left and middle panels of Supplementary Figure 4 represented by the red dots with $N_r=10,N_s=24 $,
\textit{i.e.}, a grid of $10\times25=250$ points.
The grid over $r$ builds concentric ellipses with the same aspect ratio and
$s$ angularly goes along each ellipse in steps of $\pi/N_s$.

The intrinsic parameters we seek should be chosen such that the corresponding values for $\{\chiPerp,\chiPar\}$ yield points on the elliptical grid, \textit{i.e.}, the red dots in the left panel of
Supplementary Figure 4. We start by finding the nearest point
from the set of 50,000 points (blue dots in the left panel)
to each grid point (red dot). For the $k^\text{th}$
grid point with coordinates $\{x_k,y_k\}$, we find the nearest blue dot with coordinates
$\{x_k^n,y_k^n\}$ generated from the intrinsic parameters
$\{q^n,\bar{S}_{1}^n,\bar{S}_{2}^n,\theta_{1}^n,\theta_{2}^n,\Delta\phi^n\}$.
We use these values as initial guesses in a rootfinding algorithm that translates to solving the following system
\begin{subequations}
  \begin{align}
    x_k & - \chiPerp(q,\bar{S}_1,\bar{S}_2,\theta_1,\theta_2,\Delta\phi) = 0,   \label{eq:rootfind_eq1} \\
    y_k & - \chiPar(q,\bar{S}_1,\bar{S}_2,\theta_1,\theta_2) = 0
    \label{eq:rootfind_eq2}
  \end{align}
\end{subequations}
with the caveat that the used $q$ values are consistent with our aforementioned $\eta_j$ grid.

As this is numerical root finding, we replace the right hand sides of equations~(\ref{eq:rootfind_eq1}, \ref{eq:rootfind_eq2}) with a threshold of $10^{-12}$. We perform this root finding procedure for every single elliptical grid point.
The end result is shown in the middle panel of
Supplementary Figure 4, where we place over each red dot
a faint blue dot representing the grid points that our algorithm finds.
On average, each numerically determined grid point is offset by $\le 10^{-12}$ from
the exact grid (red) point. For a grid of 250 points, this amounts a total grid offset of
$\lesssim 3\times 10^{-9}$. We actually find this number to  be $1.5\times 10^{-8}$
for the elliptical $\{\chiperp,\chieff\}$ grid of Supplementary Figure 4
because we had to relax our strict tolerance from $10^{-12}$ to $10^{-10}$
for certain grid points to speed up the procedure.
As we show further below, a grid offset of $\sim 10^{-8}$ is
much smaller than the average fit unfaithfulness that we obtain, $\sim \ord(10^{-2})$,
thus completely acceptable.

We repeated the same procedure to also obtain an elliptical grid
in the $\chiperp$-$\chipar$ plane. In the interest of expediency, we used a tolerance of $10^{-8}$
resulting in an overall grid offset of $5\times 10^{-6}$.
Let us add that a few of the intrinsic coordinates for the grid points exceed \NRsur's training limit
of $\bar{S}_i=0.8$ for spin magnitudes, but only by $\sim0.01$ which is not severe.

As is well known, rectangular domains are often best suited for constructing fits to data, therefore, we go one step further and transform the elliptical coordinates into a rectangular ones via
\begin{subequations}
  \begin{align}
    x & = X\, A \sinh \mu \cos\left( Y \right), \label{eq:rect_grid_X} \\
    y & =X\, A \cosh \mu \sin\left(Y \right)\label{eq:rect_grid_Y} ,
  \end{align}
\end{subequations}
where $X\in [0,1]$ and $Y\in [-\pi/2,\pi/2]$.
Correspondingly, we have the following inverse relations
\begin{subequations}
  \begin{align}
    X & = \f{1}{A}\ \text{csch}\mu\, \text{sech}\mu\, \sqrt{x^2\cosh^2\mu+y^2\sinh^2\mu} \label{eq:X_of_xy}, \\
    Y & = \tan^{-1}\left(\f{y\tanh \mu}{x}\right) \label{eq:Y_of_xy}.
  \end{align}
\end{subequations}
Comparing equations~\eqref{eq:ellipse_grid_x} with \eqref{eq:rect_grid_X}, and
\eqref{eq:ellipse_grid_y} with \eqref{eq:rect_grid_Y} gives the $N_r\times N_s$
rectangular grid $\{X_r,Y_s\}$ with $r=1,\ldots, N_r, s=0,\ldots N_s$, which we show in the right panel of Supplementary Figure 4.
Overall, we have the following four dimensional grid for the fitting: $\{X_r, Y_s, \eta_j,\Mtot_k\}$
with $j=1,\ldots, 5$ and $k=1,2,3$.
As a final step, we introduce the rescaled variables $Z=4\eta, V=\Mtot/(75\Msun)$.

After much trial and error, we settled on the following fitting function
\be
\fit(X,Y,Z,V) = \sum_{i=0}^{n_i}\sum_{j=0}^{n_j}
\f{\sum_{k=0}^{n_k}\sum_{l=0}^{1} c_{ijkl} \ Z^k V^l}
{\sum_{k=0}^{n_k}\sum_{l=2}^{3} \vert c_{ijkl}\vert \ Z^k V^{l-2}} \ X^i Y^j \ . \label{eq:fit_PadePade_XYZV}
\ee
We chose this particular form to better curb the fit's extrapolation behaviour in parts of the $\{Z,V\}$
(mass ratio, total mass) space outside of the training region $ Z<0.64 \ (\eta<0.16)$
and $V<1 \cup V>2$ corresponding to $\Mtot<75\Msun\cup \Mtot>150\Msun$.
We use two dimensional polynomials in the $\{X,Y\}$ subspace of the fit training domain because,
as a result of our elliptical grid design, only rare combinations of intrinsic parameters yield points just outside our outermost ellipse.
The values of $\{n_i,n_j,n_k\}$ in the triple summation of equation~\eqref{eq:fit_PadePade_XYZV} are chosen such that we have at most roughly
the same number of fit parameters as the total
number of grid points used in the $\{\chiPerp,\chiPar,\eta\}$ subspace, which in the case of Supplementary Figure 4, for example, 
is $10\times 25=250$.
Note that in the denominator of equation~\eqref{eq:fit_PadePade_XYZV},
we take the absolute value of the fit coefficients $c_{ijkl}$
to ensure that there are no singularities. We also set $c_{ij02}=1$, which is
the leading term in the denominator, a standard choice for Pad\'{e} type fits.
Our general procedure is as follows: \\
\begin{inparaenum}[(i)]
  \item start with a large ensemble of intrinsic parameters
  $\{q_i,\mathbf{S}_{1,i},\mathbf{S}_{2,i}\}$ for $i=1,\ldots, \ord(10^4)$.\\
  \item Impose an elliptical grid of size $N=N_r\times (N_s+1)$ with the grid
  coordinates given by equations~\eqref{eq:ellipse_grid_x} and \eqref{eq:ellipse_grid_y}.\\
  \item Determine the set of intrinsic parameters $\{q_I,\mathbf{S}_{1,I},\mathbf{S}_{2,I}\}$
  yielding this grid to some tolerance, \textit{e.g.}, $10^{-12}$.\\
  \item Compute the mismatches $\mismatch_{K,L}$ of $L$ models to \NRsur{} for the set
  $\{\Mtot_K,q_K,\mathbf{S}_{1,K},\mathbf{S}_{2,K}\}$ where $K=3I$ for the three distinct values of $\Mtot$ that we use.\\
  \item Transform to the rectangular grid $\{X_K,Y_K,Z_K,V_K\}$.\\
  \item For each model $L$, perform the fitting to the set $\{X_K,Y_K,Z_K,V_K,\log_{10}\mismatch_{K,L}\}$ using
  \textsc{Mathematica}'s \texttt{NonlinearModelFit} function and
  store the coefficients $c_{ijkl,L}$ of equation~\eqref{eq:fit_PadePade_XYZV}.
\end{inparaenum}

We start our fit optimization routine with $\{ n_i,n_j,n_k\}=\{4,3,2\}$ and generate fits
up to some $\{n_i^\text{max}, n_j^\text{max}, n_k^\text{max}\}$ that ensures that the total number
of fit parameters is $\approx N$. The choice of $\{4,3,2\}$ yields 160 fit parameters. Smaller values
of $\{ n_i,n_j,n_k\} $ result in fewer than 100 fitting coefficients and leads to underfitting for training grids of size $\gtrsim \ord(200)$
which, as we explain below, is the grid size that we adopt.
Our routine picks as the final fit the one for which the values of $\{n_i,n_j,n_k\}$ in equation~\eqref{eq:fit_PadePade_XYZV}
yield the lowest relative difference with respect to the \emph{training} data set.
For this purpose, we define the relative difference between the data and the fit at the $k^\text{th}$ point
\be
\Delta^k_\text{rel}:=1-\f{\fit(X_k,Y_k,Z_k,V_k)}{\log_{10}\mismatch_k},\label{eq:rel_diff_at_each_pt}
\ee
to introduce two quantities to gauge fit quality during training. The first is the $l^2$-norm 
of $\Delta^k_\text{rel}$ between the fit and the data normalized by the length of the vector
\be
\Delta_\text{rel}^{(1)} := \f{1}{3N} \sqrt{\sum_{k=1}^{3N} \left\vert \Delta^k_\text{rel} \right\vert^2  }\label{eq:rel_diff1},
\ee
and the second is the signed average relative difference
\be
\Delta_\text{rel}^{(2)} := \f{1}{3N} \sum_{k=1}^{3N}  (\Delta^k_\text{rel}) \label{eq:rel_diff2}
\ee
which tells us whether the fit globally over or underestimates the data.

We pick the values for $\{n_i,n_j,n_k\}$ that simultaneously minimize both of the above relative differences. These
relative differences are the most important fit attributes for this work
as we must robustly predict the mismatches to NR in order to appropriately select which model to use at a given point in
the parameter space.
If more than one set of values for $\{n_i, n_j, n_k\}$ are returned, we opt for the set which yields a
reduced chi square ($\chi^2/\text{DoF}$) closest to unity.
Once the fit training is complete via the above optimization of $\{n_i,n_j,n_k\}$, we check fit performance
over an appropriate verification set which we discuss further below.

The question of training grid resolution can only be answered after setting a target fit unfaithfulness 
threshold. Here, we aim for $\Delta_\text{rel}^{\vert\text{av}\vert}\approx 0.05$ for each fit,
where
\be
\Delta_\text{rel}^{\vert\text{av}\vert}:=\f{1}{3N} \sum_{k=1}^{3N} \vert\Delta^k_\text{rel}\vert \label{eq:rel_diff_abs}
\ee
is the average absolute relative disagreement between the fit and the verification data.
With the above threshold established, we set out to determine whether or not the $\{x,y\}$ training grid of size $10\times25$ suffices.
First, we downsample this grid to create coarser grids of dimension $10\times 13,\, 5\times 13,$ and $5\times 7$
and compute $\Delta_\text{rel}^{\vert\text{av}\vert}$ for each with respect to the original
verification set (of size 250). 
As we gradually increase the grid size from $5\times 7$ to $10\times 25$,
we observe $\Delta_\text{rel}^{\vert\text{av}\vert}$ decreasing from $\approx 0.10$ to $\lesssim 0.05$ for the fits listed in 
Supplementary Table 2.
For example, the fit used to make Supplementary Figure 5 yields
$\Delta_\text{rel}^{\vert\text{av}\vert}=0.048$.
Increasing the grid size to $\ord(1000)$ elements should further reduce $\Delta_\text{rel}^{\vert\text{av}\vert}$. However, this quickly turns into a problem of diminishing returns
given that it would
take one month to generate this mismatch data using 128 CPUs on this grid.

Furthermore, an inspection of the structure of the mismatch data revealed that an elliptical grid with $\ord(10)$ points along the radial 
direction and $\ord(20)$ points along the azimuthal direction suffices to capture the dominant trends in the data
at the level of fit unfaithfulness that we seek, \textit{i.e.},
$\Delta_\text{rel}^{\vert\text{av}\vert}\approx 0.05$.
Thus, our aforementioned grid of dimensions $10\times 25$ has sufficient resolution.
We leave fit improvements to future work which we have already begun undertaking.

In Supplementary Figure 5, we show a contour plot of the unfaithfulness of the fit for the $\log_{10}$ of the \NRsur-\modelname{SEOBNRv5PHM} mismatches to the verification data set.
Due to the computational burden of obtaining the mismatches, 
we use data with $y=\chipar\ (\chieff)$ to verify the data trained with $\chieff\ (\chipar) $.
This results in verification sets that are
the same size as the training sets, so ours is rather a harsh verification test.
The contours represent the absolute value of the relative difference between the fit and the data.
The fit is trained over the $y=\chipar$ set (black dots) which, by design, trace concentric prolate ellipses in the $\{x,y\}=\{\chiperp,\chipar\}$ plane.
The white dots mark the $\{x,y\}$ coordinates of the verification data.
From the figure, we see that in a large portion of the space, the relative difference is $0.05$ or less.
Note that this quantity is not
$\Delta_\text{rel}^{(2)}$ applied to the verification set, but rather the absolute value of the summand in equation~\eqref{eq:rel_diff2}.
The fits for the \NRsur-\modelname{IMRPhenomTPHM} and the \NRsur-\modelname{IMRPhenomXPHM} mismatches
also yield similar level of agreement as do the fits trained by the $y=\chieff$ set.

We summarize these results and provide additional metrics for all the fits in Supplementary Table 2 where we see that the average relative distance \eqref{eq:rel_diff1}
between each fit and the corresponding verification data is always less than $ 4\times 10^{-3}$
and the average relative difference \eqref{eq:rel_diff2} has magnitude less than $0.02$.
Interestingly, we observe that $\Delta_\text{rel}^{(2),\text{ver}}$ is negative for most fits
indicating that the fits are slightly overestimating the data.
The value of $1-\tilde{R}^2$ is $\lesssim 0.01$
for all our fits, where $\tilde{R}^2$ is the reduced $R$ square
and for most cases we observe $\chi^2/\text{DoF} \approx \ord(1)$.
The cases for which this quantity is about an order of magnitude lower stem from the
fact that we overestimate our errors bars.
Recall that these are actually the standard deviations of an ensemble of mismatches
(over a grid of certain extrinsic parameters per a given set of intrinsic parameters)
whose average we take to be our individual data points.

As an additional check of the fits, we investigate their behavior in the extrapolation region
corresponding to $\eta<0.16\ (\text{i.e., }q<1/4)$, $\bar{S}_{1,2}> 0.8$ and $\Mtot < 75\Msun \cup \Mtot > 150\Msun$.
Recall that we chose the particular form of equation~\eqref{eq:fit_PadePade_XYZV} for the fitting function
to better control unwanted extrapolation behavior such as blow-ups common to
polynomial fitting.
Specifically, the Pad\'{e} type dependence on $\eta$ and $\Mtot$
was adopted so that the fits would not produce any nonsensical results such as
$\log_{10}\mismatch > 0$ in regions of the $\{\eta,\Mtot\}$ space quite distant
from the training (interpolation) regime. On the other hand, since the relevant 2D cut of
the training region covers most of the $\{\chiPerp, \chiPar\}$ space, polynomial extrapolation
should not cause issues.

We illustrate all of this in Supplementary Figure 6, where
we plot the fit in equation~\eqref{eq:fit_PadePade_XYZV} to the $\log_{10}$ of the
\NRsur-\modelname{SEOBNRv5PHM} mismatches as a function of $\Mtot$,
evaluated at various extrapolated values of $\{\eta,\bar{S}_1=\bar{S}_2\}$.
The blue ellipse in the $\chiperp$-$\chieff$ plane traces the values
$\{\eta,\bar{S}_{1,2}\}=\{0.139,0.85\}\ (q=1/5)$ with other intrinsic parameters chosen accordingly.
Similarly, the red ellipse traces the
$\{\eta,\bar{S}_{1,2}\}=\{0.122,0.9\}\ (q=1/6)$ set and the orange ellipse the edge of the fit training region
with $\{\eta,\bar{S}_{1,2}\}=\{0.16,0.8\}\ (q=1/4)$, which was already shown in Supplementary Figure 4.
The blue, red and orange dots mark the positions of seven cases along each corresponding
ellipse in angular steps of $\pi/6$.
An inset pointing to each dot displays the plot of the fit from $\Mtot=50\Msun$ to $200\Msun$, but at each separate elliptical coordinate, hence the blue, red, orange colored curves.
The shaded gray region in each inset marks the training range of $\Mtot\in[75,150]\Msun$
for the fits, only actually relevant to the orange curves as the blue and the
red are, by definition, outside the training region.
Thanks to the specific functional form of the fit,
the extrapolation does not exhibit any pathologies.
Additionally, we note that the blue curves mostly lay between the orange and the red ones
as we would expect. We should, however, caution that we are merely demonstrating that
the extrapolation is not pathological. This does not mean that the fits are expected be faithful
to the data in this regime. As such we recommend their use in the regime $q\ge 1/5, \bar{S}_i \le 0.85$.

As a further test of the fit's performance in its extrapolation regime, we performed two more parameter recoveries of an NR simulation, \texttt{SXS:BBH:1156}, with the injected value for $\Mtot$
set to $75\Msun$ and ${100\Msun}$, and $q=0.228$ placing the former ``squarely''
outside the $M>75\Msun, q\ge 1/4$ training regime of our fits.
This simulation also has $\vert \mathbf{S}_{2,\perp}\vert/m_2^2\approx 0.76$ while the primary has negligible planar spin.
We show the results of our method applied to this Bayesian inference analysis in Supplementary Figure 3 with the 2D posteriors from the $\Mtot=75\Msun$ ($100\Msun$) injections plotted in the top (bottom) panels.
We can see that in both analyses, our method recovers the injected values for $m_1,m_2$ and the effective spins very well with the majority of the samples clustered near the injected values.
Such a good recovery of the masses, hence the mass ratio, for the $\Mtot=75\Msun$ injection is an indirect testament to the robustness of the
fit \eqref{eq:fit_PadePade_XYZV}, especially since slightly more than half the total mass posteriors
for this particular Bayesian analysis happen to be less than $75\Msun$.

We conclude this section by briefly returning to two issues: the first regarding the fact that the  fitting variables
$\{\eta,\chiPerp,\chiPar\}$ used to construct the fit are not fully independent of each other, as
each one is a function of the mass ratio $q$, and the second regarding the choice of power spectral density used when calculating the mismatch and hence the fits. For the first issue, as we explained already, the $q$-scaled spin projections yield more faithful fits
to the data.
As a check, we computed the correlation coefficients $\mathcal{C}_{mn}$
between the above parameters of the training sets $K_1 (\chiPar=\chieff)$,
$K_2 (\chiPar=\chipar)$. For $K_1$, we obtain
$\mathcal{C}_{\chiperp\text{-}\chieff} = 0.09$, $\mathcal{C}_{\eta\text{-}\chieff} = 0.02$
and $\mathcal{C}_{\eta\text{-}\chiperp} = -0.315$.
For $K_2$, we have $\mathcal{C}_{\chiperp\text{-}\chipar} = -0.1$, $\mathcal{C}_{\eta\text{-}\chiperp} = 0.03$
and $\mathcal{C}_{\eta\text{-}\chipar} = 0.216$.
For both fit-training parameter sets, we have either uncorrelated pairings of fit variables
or weakly correlated pairings.
The fact that $\vert\mathcal{C}_{\eta\text{-}\chipar}\vert$ of set $K_2$ is less than
$\vert\mathcal{C}_{\eta\text{-}\chiperp}\vert$ of set $K_1$ may partly explain
why we observe a slightly better performance from the fits constructed from $K_2$
as indicated in Supplementary Table 2.

For the second issue, as already explained, we use the Advanced LIGO's design sensitivity\cite{dcc:T2200043} when computing the mismatch and this will subtly change when the profile of the power spectral density is varied. If this method were to be used during live observing run periods, where the power spectral density is changing on a day-to-day basis owing to noise artefacts in the GW strain data, we would suggest using a harmonic average power spectral density estimated from engineering run data to calculate mismatches, as is commonly done in GW search pipelines~\cite{DalCanton:2017ala}. The fit would then be reconstructed prior to each GW observing run.

\section*{Data Availability}
The aligned-spin and generic-spin match interpolants as well as the posterior samples from the analyses performed in this work are available at \href{https://icg-gravwaves.github.io/incorporating_model_ uncertainty_into_Bayesian_inference}{https://icg-gravwaves.github.io/incorporating\_model\_ uncertainty\_into\_Bayesian\_inference}

\section*{Code Availability}
Python scripts detailing our modifications to {\texttt{Bilby}}\cite{Ashton:2018jfp} are available at \href{https://icg-gravwaves.github.io/incorporating_model_ uncertainty_into_Bayesian_inference}{https://icg-gravwaves.github.io/incorporating\_model\_ uncertainty\_into\_Bayesian\_inference}

\section*{Acknowledgements}
We would like to thank Anjali Yelikar for comments during the LIGO-Virgo-KAGRA internal review, as well as Christopher Berry, Alexandre G\"{o}ttel, Maximiliano Isi, Lucy Thomas, Michael Williams and Aaron Zimmerman for discussions during our presentation to the LIGO-Virgo-KAGRA collaboration. We are also grateful to Mark Hannam and Laura Nuttall for discussions throughout this project. We thank Alvin Chua and the California Institute of Technology for hosting CH and SA in March 2024, giving the authors time to discuss and develop this project. CH thanks the UKRI Future Leaders Fellowship for support through the grant MR/T01881X/1, SA and JMU acknowledge support from the University College Dublin Ad Astra Fellowship, and JT acknowledges support from the NASA LISA Preparatory Science grant 20-LPS20-0005. This work used the computational resources provided by the ICG, SEPNet and the University of Portsmouth, supported by STFC grant ST/N000064. This research made use of data, software and/or web tools obtained from the Gravitational Wave Open Science Center (\href{https://www.gw-openscience.org}{https://www.gw-openscience.org}), a service of LIGO Laboratory, the LIGO Scientific Collaboration and the Virgo Collaboration. LIGO is funded by the U.S. National Science Foundation. Virgo is funded by the French Centre National de Recherche Scientifique (CNRS), the Italian Istituto Nazionale della Fisica Nucleare (INFN) and the Dutch Nikhef, with contributions by Polish and Hungarian institutes. This material is based upon work supported by NSF's LIGO Laboratory which is a major facility fully funded by the National Science Foundation.

\section*{Author contributions}
CH conceptualised the idea of sampling over multiple models, with priors dictated by their mismatch to numerical relativity simulations, as a method to incorporate model uncertainty into GW bayesian inference. SA and JT developed the idea of building mismatch interpolants for this application. SA and JT initiated and formed the project team. CH implemented the method presented here into {\texttt{Bilby}}\cite{Ashton:2018jfp} and performed all parameter estimation analyses. JT generated the aligned-spin mismatches and the aligned-spin interpolant. SA constructed the generic-spin interpolant. 
CH and JT investigated choices for the model conditional prior. JMU generated the majority of generic-spin mismatches, with CH and SA generating a subset. All authors contributed to the interpretation of the results and wrote the paper.

\section*{Competing interests}
The authors declare no competing interests.

\bibliography{refs}

\section*{Supplementary Material}
\label{Sec:supplement}

\renewcommand{\figurename}{Supplementary Figure}
\setcounter{figure}{0}
\renewcommand{\tablename}{Supplementary Table}
\setcounter{table}{0}

\begin{figure*}[t!]
  \centering
  \includegraphics[width=0.48\textwidth]{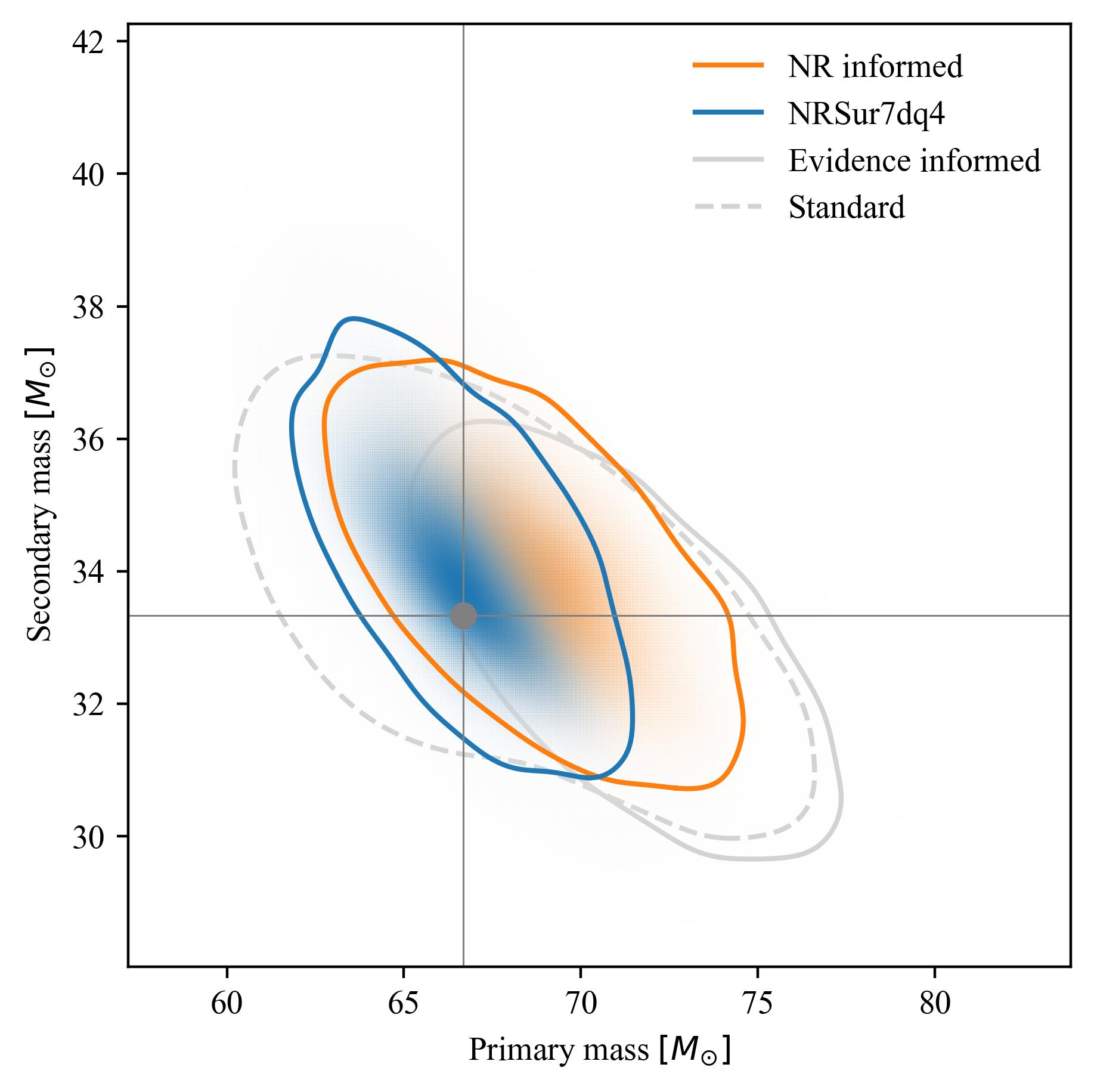}
  \includegraphics[width=0.48\textwidth]{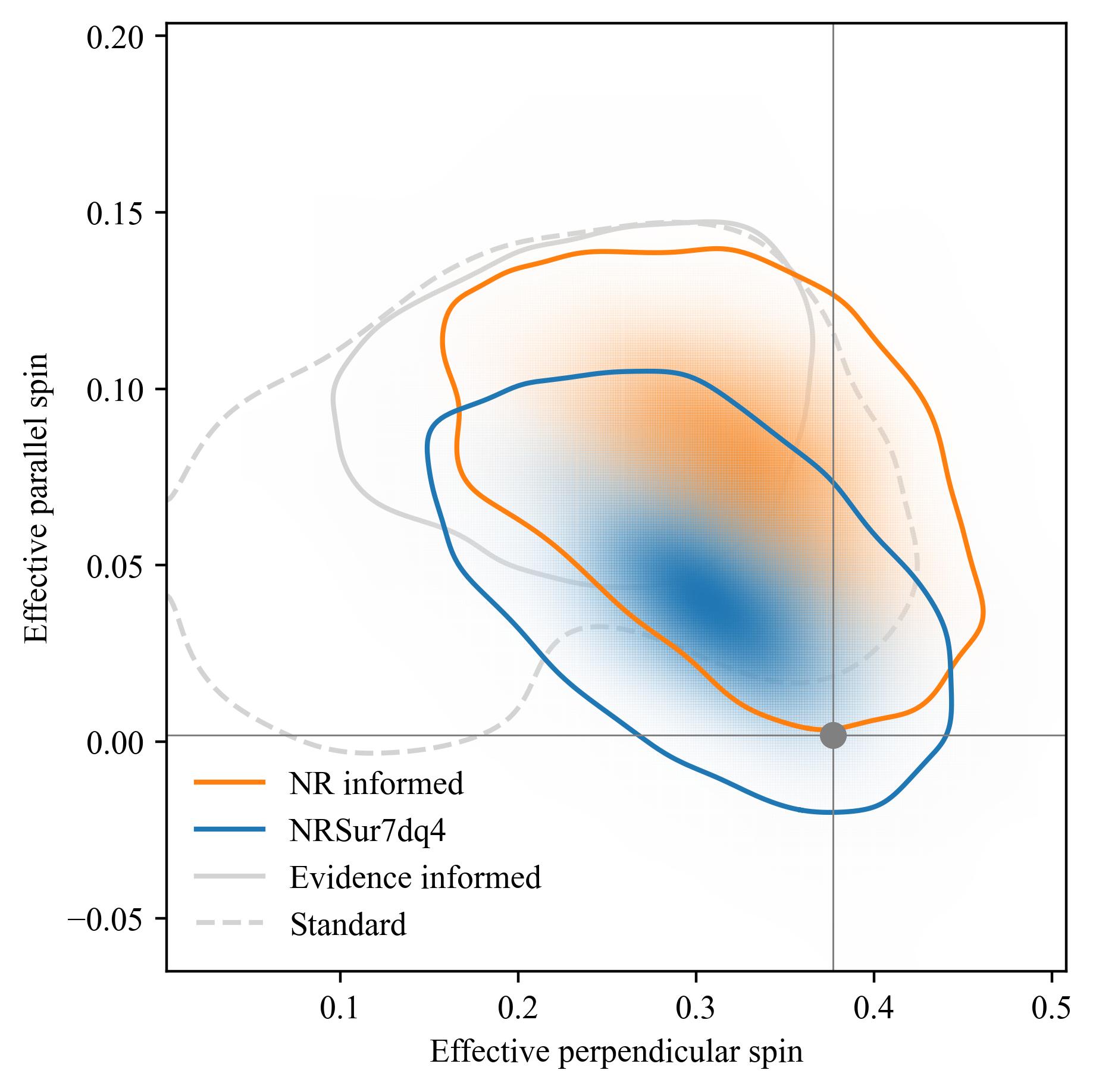}
  \caption{Two-dimensional posterior probabilities obtained in our analysis of the 
  {\texttt{SXS:BBH:0926}} numerical relativity simulation. The left panel shows the measurement of the primary and secondary mass of the binary, and the right panel shows the inferred effective parallel and perpendicular spin components (as defined in equation~\ref{eq:chiperp}, \ref{eq:chi_par}). The contours represent 90\% credible intervals and the grey cross hairs indicate the true value. The Evidence-informed and Standard analyses are set to a lighter color than in other Figures to highlight the comparison between NR informed and \modelname{NRSur7dq4}.}
  \label{fig:nrsur_comparison_plot}
\end{figure*}

\begin{figure*}[t!]
  \centering
  \includegraphics[width=0.48\textwidth]{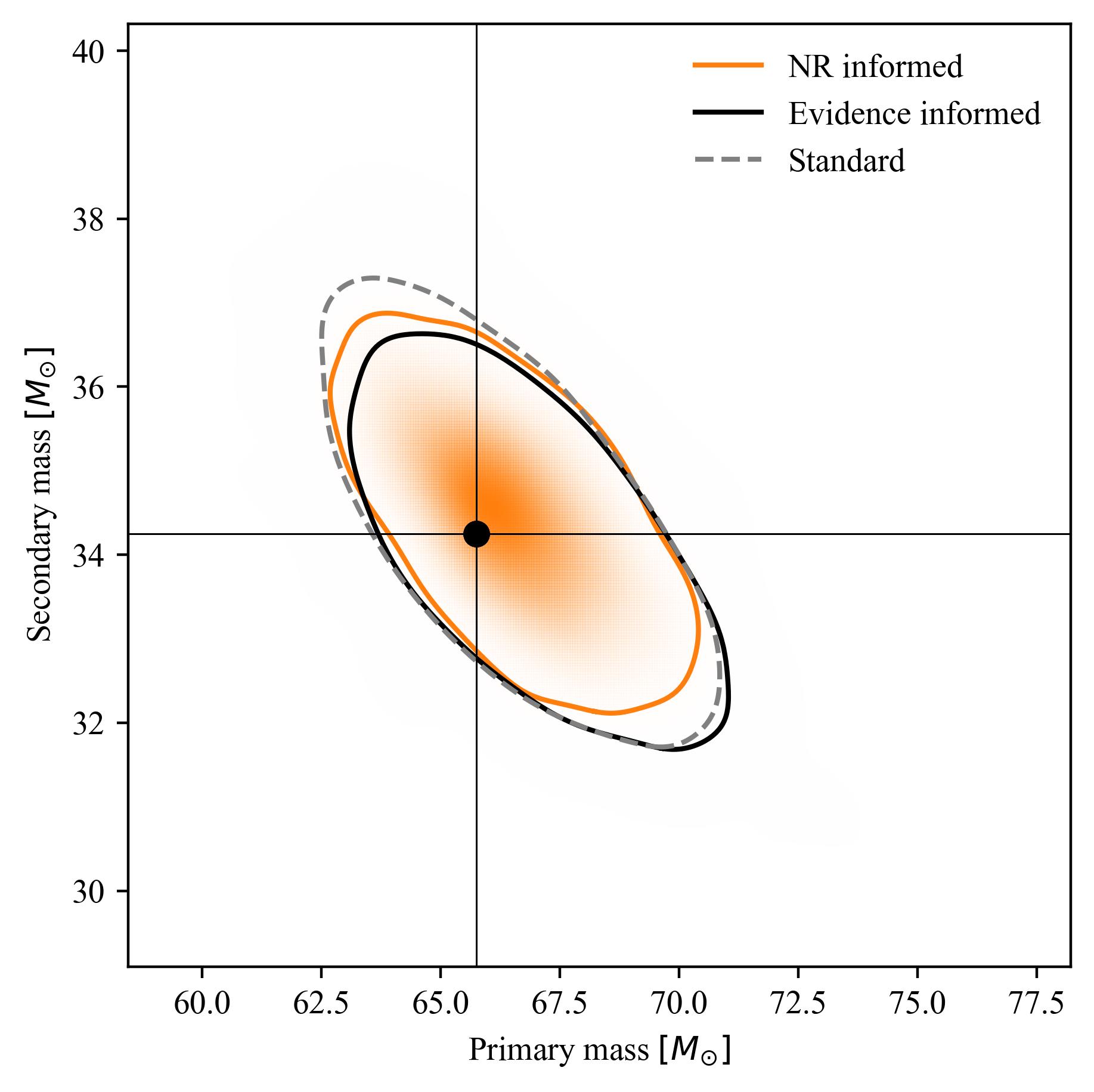}
  \includegraphics[width=0.48\textwidth]{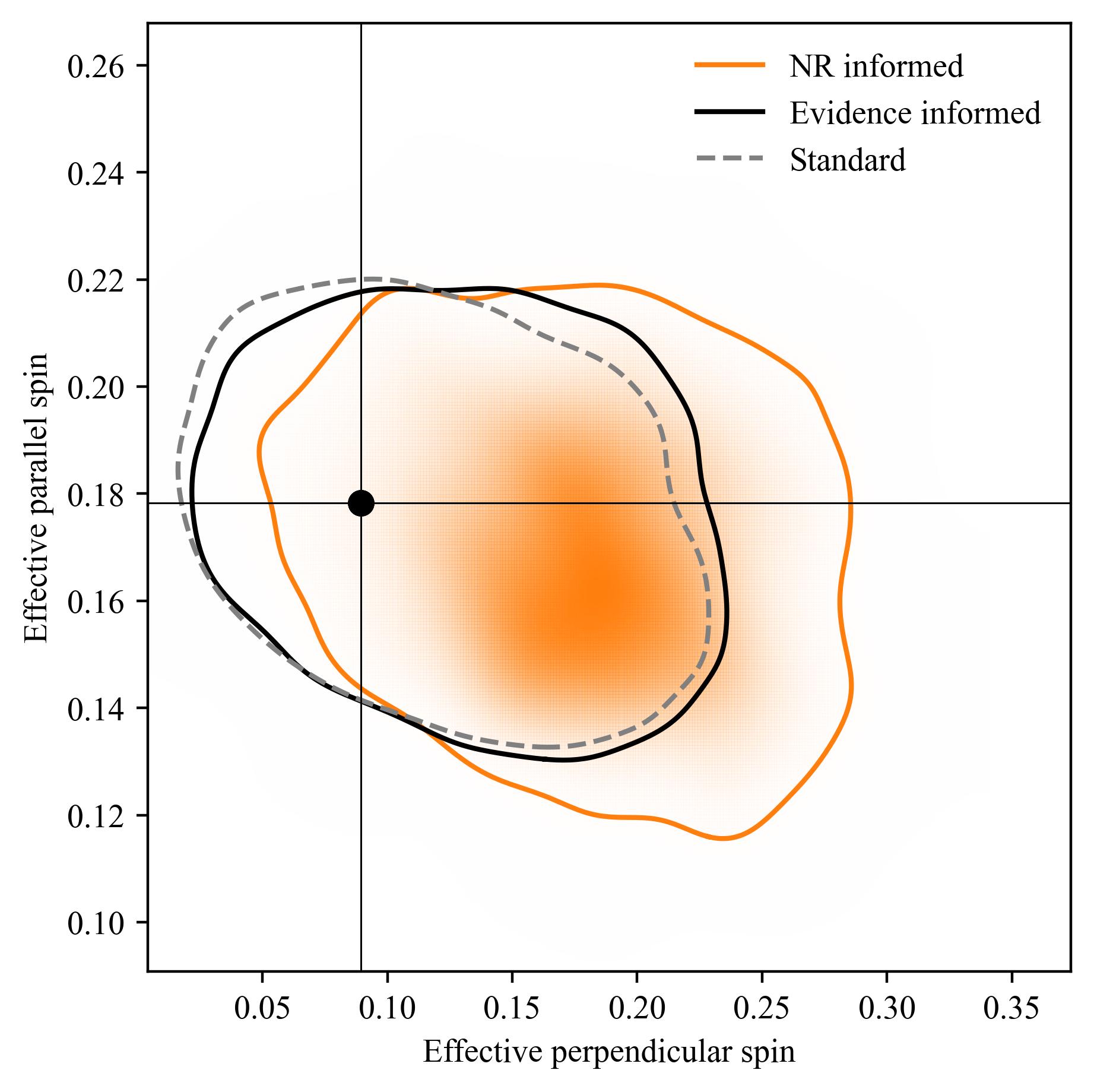}
  \caption{Two-dimensional posterior probabilities obtained in our analysis of the {\texttt{SXS:BBH:0143}} numerical relativity simulation. The left panel shows the measurement of the primary and secondary mass of the binary, and the right panel shows the inferred effective parallel and perpendicular spin components (as defined in equation~\ref{eq:chiperp}, \ref{eq:chi_par}). The contours represent 90\% credible intervals and the black cross hairs indicate the true value.}
  \label{fig:0143_comparison_plot}
\end{figure*}

To demonstrate the improved accuracy of the NR informed approach compared
to Evidence informed and Standard algorithms, we compared the posterior
distributions obtained with each method against those obtained with 
{\modelname{NRSur7dq4}}; the leading generic spin numerical relativity surrogate model.
We only considered the {\texttt{SXS:BBH:0926}} numerical relativity simulation since it was
the main focus of this work.

In Supplementary Figure~\ref{fig:nrsur_comparison_plot} we see that our NR informed approach
infers a posterior that significantly overlaps with {\modelname{NRSur7dq4}}. For the inference
of the binary masses we see that NR informed has support for slightly larger primary masses,
but the secondary mass remains comparable with {\modelname{NRSur7dq4}}. Similarly, when
considering the inferred component spins, we see that our NR informed approach infers a
comparable marginalized one-dimensional posterior for the effective perpendicular spin,
with slightly more support for larger effective parallel spin. Importantly, we see that the 
Evidence informed and Standard approaches fail to capture the true value. Specifically, the
Evidence informed result has low overlap with the {\modelname{NRSur7dq4}} 
result and the Standard result unnecessarily inflates the uncertainty for the effective 
perpendicular spin. Interestingly, while both our NR informed approach and
{\modelname{NRSur7dq4}} capture the true effective spins of the binary with their two-dimensional 90\%
credible intervals, the majority of posterior support is to lower effective perpendicular and larger
effective parallel spins. In general, we found that our NR informed approach obtained the most
statistically similar one-dimensional posterior probability distributions to the surrogate out of
the methods considered in this work.

We additionally analysed the {\texttt{SXS:BBH:0143}}\cite{Blackman:2017pcm,Boyle:2019kee} and {\texttt{SXS:BBH:1156}}\cite{Blackman:2017pcm,Boyle:2019kee} numerical relativity simulations produced by the SXS collaboration with our NR informed approach, as well as the Evidence informed and Standard algorithms. For the case of {\texttt{SXS:BBH:0143}}, we found largely overlapping posteriors between methods, with most of the one-dimensional marginalized 90\% confidence intervals containing the true value. For the case of {\texttt{SXS:BBH:1156}}, we found that our method outperforms the Standard and Evidence informed analyses despite being in the extrapolation regime of our technique. Here, we provide further details.

\begin{figure*}[t!]
  \centering
  \includegraphics[width=0.48\textwidth]{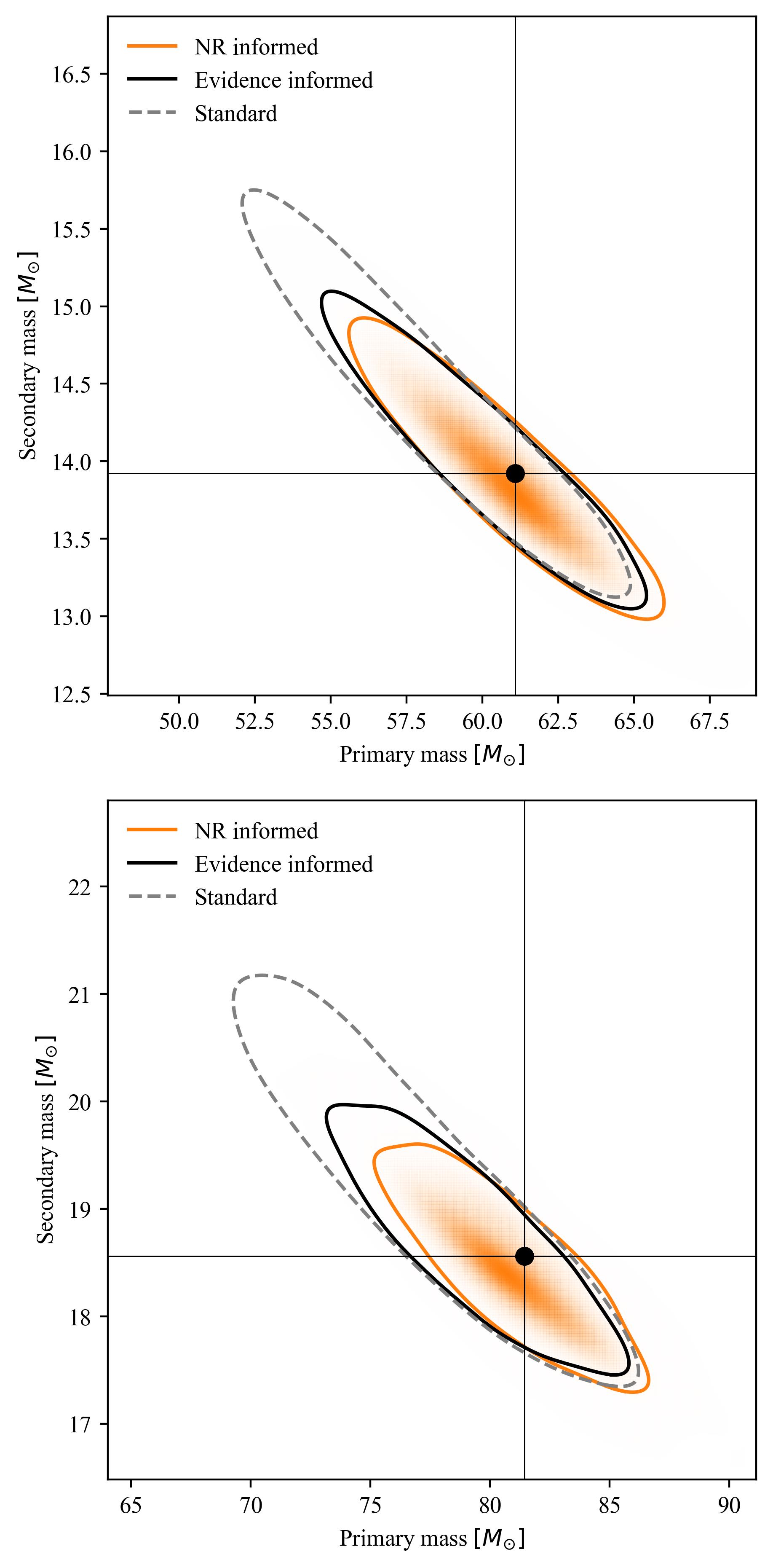}
  \includegraphics[width=0.48\textwidth]{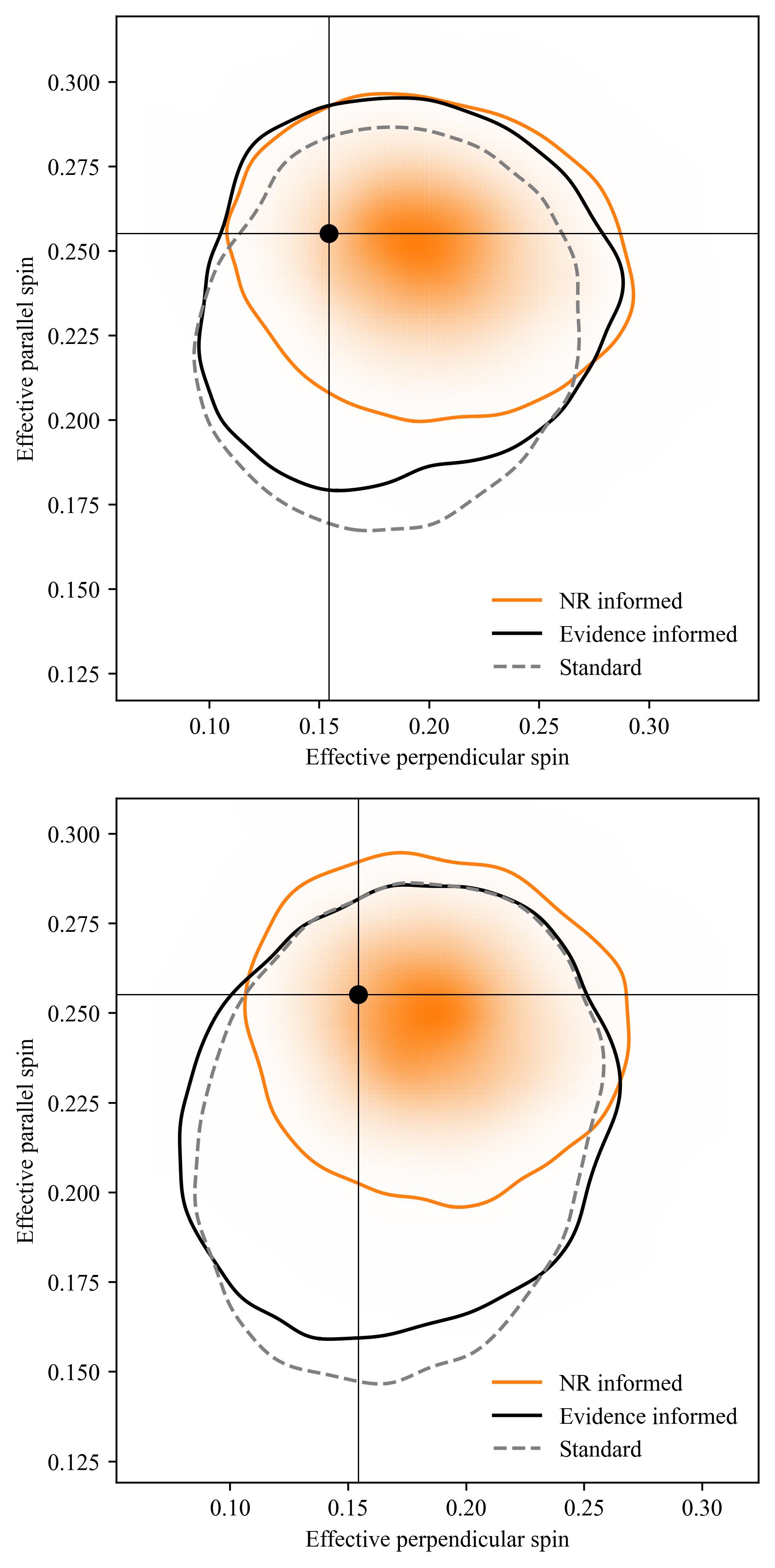}
  \caption{Supplementary Figure 3: Two-dimensional posterior probabilities obtained in our analysis of the {\texttt{SXS:BBH:1156}} numerical relativity simulation. The left column shows the measurement of the primary and secondary mass of the binary, and the right column shows the inferred effective parallel and perpendicular spin components (as defined in equations~(\ref{eq:chiperp}, \ref{eq:chi_par})). The top row shows a low total mass injection of $M = 75\, M_{\odot}$ and the bottom row shows a high total mass injection of $M = 100\, M_{\odot}$. The contours represent 90\% credible intervals and the black cross hairs indicate the true value.}
  \label{fig:1156_comparison_plot}
\end{figure*}

In Supplementary Figure~\ref{fig:0143_comparison_plot} we show the results obtained with our method, NR informed, vs. the Evidence informed and Standard analyses for the parameter recovery
of the {\texttt{SXS:BBH:0143}} numerical relativity simulation. We see that in general, all methods give largely overlapping posteriors. When inspecting the one-dimensional marginalized posterior distributions, provided in the associated data release, the biggest disagreement is seen in the inferred secondary spin magnitude of the binary: our method prefers a rapidly spinning black hole, while the other methods are largely uninformative with a spin $\sim 0.25$. The reason is because our method primarily chooses the {\modelname{SEOBNRv5PHM}} model, while the Evidence informed result is primarily using {\modelname{IMRPhenomTPHM}}. In this region of the parameter space we find that {\modelname{SEOBNRv5PHM}} is the most accurate model, hence why it is favoured in our method: we find that {\modelname{SEOBNRv5PHM}} is $\sim 1.4\times$ more accurate than {\modelname{IMRPhenomTPHM}} and $\sim 1.8\times$ more accurate than {\modelname{IMRPhenomXPHM}}. Given that the individual spin components are difficult to measure at present-day detector sensitivities\cite{Purrer:2015nkh}, it is not surprising that the biggest difference between methods is seen for the secondary spin. Since the secondary spin contribution is also suppressed for asymmetric mass ratio binaries, it is possible that the mismatch interpolant is missing information, leading to poorly chosen models in this region of the parameter space. A mismatch interpolant that is a function of the individual spin components may solve this problem, but it would require a significantly larger mismatch dataset. We leave this to future work. It could also be independent of the interpolant: the interpolant could be accurately describing the mismatch, but {\modelname{SEOBNRv5PHM}} may not be accurate enough to avoid parameter biases, despite being the most accurate model of those considered in this region of the parameter space.

In Supplementary Figure~\ref{fig:1156_comparison_plot} we compare results for the {\texttt{SXS:BBH:1156}} numerical relativity simulation. Although all methods infer the injected value within 90\% credible interval, we see that our method outperforms the others, especially seen for the high total mass ($M = 100\, M_{\odot}$) injection: only our method contains the injected spin within the 40\% confidence interval. When inspecting the one-dimensional marginalized posterior distributions, provided in the associated data release, we consistently see a more accurate inference of the binary parameters with our method. As with {\texttt{SXS:BBH:0143}}, we see the worst performance for the inferred spin of the smaller black hole in the binary. Although the injected value is contained within the 90\% confidence interval, the Evidence informed and Standard analyses capture the true value more accurately. For the low total mass ($M = 75\, M_{\odot}$) injection, we observe a better performance by all methods. We argue that this is because the GW produced from a lower mass system spends more time within the sensitive region of the GW detectors, meaning that all GW models better capture the true source parameters of the binary, as long as the primary is not strongly precessing.

\begin{table*}[t!]
  \begin{center}
    \begin{tabular}{l | c l l}
      \hline
      \hline
      Primary spin, $\chi_{1} / m_{1}$   & 0.45 & mbits \\
      Secondary spin, $\chi_{2} / m_{2}$ & 0.51 & mbits \\
      Mass ratio, $q = m_2/m_1$          & 0.89 & mbits \\
      Phase, $\phi$                      & 2.1  & mbits \\
      Polarization, $\Psi$               & 0.82 & mbits \\
      Right ascension $\alpha$           & 0.99 & mbits \\
      Declination $\delta$               & 1.3  & mbits \\
      \hline
      \hline
    \end{tabular}
    \caption{Comparison of Jensen-Shannon divergences when using the true mismatch. Jensen-Shannon divergences between posteriors obtained when using true mismatch and the mismatch interpolant for an aligned-spin injection and recovery. 
    Column one lists the parameters for which we display the Jensen-Shannon divergences of the posteriors in column two.
    These Jensen-Shannon divergences are reported in the base 2 logarithm and reported in millibits (mbits).}
    \label{tab:js_divergences}
  \end{center}
\end{table*}

\begin{figure*}[t!]
  \centering
  \includegraphics[width=0.98\textwidth]{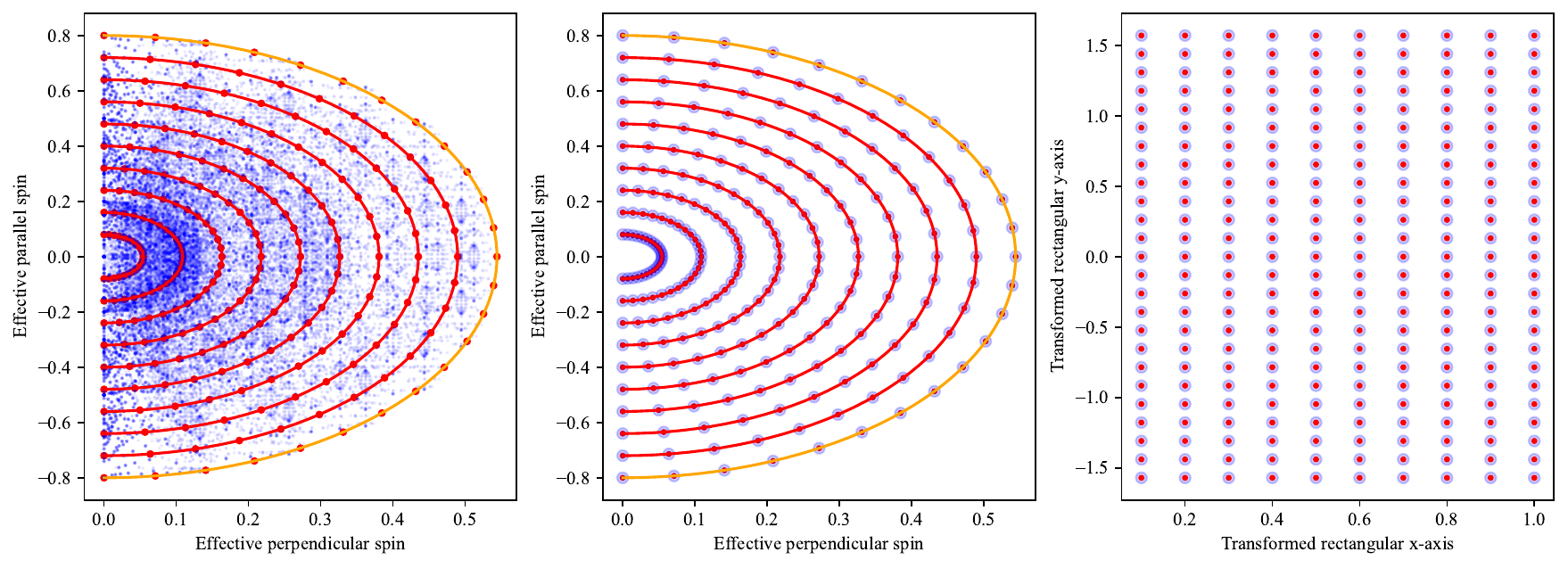}
  \caption{Construction of the two dimensional elliptical and rectangular grids. We show the construction of the two dimensional elliptical and rectangular grids for
    the $\{\chiPerp,\chiPar\}=\{\text{Effective perpendicular spin}, \chiperp, \text{Effective spin}, \chieff\}$ pairing.
    The blue dots in the left panel represent the $\sim 50000$ points obtained from a regularly spaced grid
    in the intrinsic parameter space $\{q,\bar{S}_1,\bar{S}_2,\theta_1,\theta_2,\Delta\phi\}$.
    The orange curve is the outermost prolate ellipse whose axes are determined via equation (13). The red curves are concentric ellipses retaining the same aspect ratio as the orange one.
    These are parametrized by $r,s$ in equations~(16a, 16b).
    The red and orange dots situate the elements of the $10\times 25$ elliptical grid where the azimuthal coordinate
    is sampled in steps of $\pi/24$ from $-\pi/2$ to $\pi/2$.
    Though this grid size is somewhat arbitrary, it is limited by computational resources such as the
    runtimes to generate the mismatches (see Sec. 4.5) and to construct the fits.
    The middle panel shows the same elliptical grid of the left panel overlaid, in faint blue, with the actual grid points that we determine by finding the roots of equations~(17a, 17b).
    On average, each blue dot is offset by $10^{-12}$ from the nearest, exactly positioned red dot.
    Finally, in the right panel, we show the corresponding rectangular grid.
    The red dots are exact, hence simply   given by the transformed rectangular coordinates $X=0.1,0.2,\ldots, 1, Y=-\pi/2, -\pi/2+\pi/24,\ldots, \pi/2$ and the blue dots are mapped from the blue dots of the middle panel via
    equations~(19a, 19b).}
\end{figure*}

\begin{table*}[t!]
  \centering
  \begin{tabular}{lccccccc}
    Model                                & $\chiPar$          & $\{n_i,n_j,n_k\}$   & \# Params & $\Delta_\text{rel}^{(1),\text{ver}}$ &
    $\Delta_\text{rel}^{(2),\text{ver}}$ & $1-\tilde{R}^2$    & $\chi^2/\text{DoF}$                                                                                        \\
    \hline
    \hline
    \modelname{SEOBNRv5PHM}              & $\chi_\text{eff} $ & $\{4, 3, 2\}$       & $220 $    & $0.0032 $          & $-0.013 $  & $0.0049 $ & $0.050 $ \\
   \modelname{SEOBNRv5PHM}              & $\chi_\parallel $  & $\{5, 3, 2\}$       & $264 $    & $0.0026 $                            & $0.0041 $  & $0.0052 $ & $1.3 $   \\
   \modelname{IMRPhenomTPHM}            & $\chi_\text{eff} $ & $\{5, 3, 2\}$       & $264 $    & $0.0039 $                            & $-0.012 $  & $0.012 $  & $0.99 $  \\
    \modelname{IMRPhenomTPHM}            & $\chi_\parallel $  & $\{4, 3, 2\}$       & $220 $    & $0.0039 $                            & $-0.0090 $ & $0.0077 $ & $0.076 $ \\
    \modelname{IMRPhenomXPHM-ST}         & $\chi_\text{eff} $ & $\{4, 3, 2\}$       & $220 $    & $0.0038 $                            & $-0.019 $  & $0.010 $  & $1.2 $   \\
    \modelname{IMRPhenomXPHM-ST}         & $\chi_\parallel $  & $\{4, 3, 2\}$       & $220 $    & $0.0032 $                            & $-0.0021 $ & $0.0078 $ & $0.74 $ \\
    \hline
    \hline
  \end{tabular}
  \caption{Metrics for the mismatch fitting function. Relevant metrics for the fits to the $\log_{10}$ of the mismatches between
    \NRsur{} and the models listed in the first column. The fitting function is given in
    equation (20) with the transformation from the $\{X,Y\}$ coordinates to the elliptical $\{\chiPerp,\chiPar\}$ done via equations~(19a, 19b).
    Column two lists our choice for $\chiPar$ representing whether we used $\chieff$ or $\chipar$ when
    constructing the elliptical fit training grid, \textit{e.g.}, shown for $\chieff$ in Supplementary Figure 4. Column three lists the upper limits of the triple summation
    used in equation~(20) which then determines the total number of fitting
    parameters $c_{ijkl}$ displayed in column four. Columns five and six present the values for
    the two relative difference measures of equations~(22, 23)
    applied to the verification sets.
    Finally, column seven lists $1-\tilde{R}^2$, where $\tilde{R}^2$ is the reduced $R$ square
    and column eight the chi square over the degrees of freedom.
    Overall, we see that the fits based on the $\chiPar=\chipar$ grid are slightly more faithful to the verification
    data and $\tilde{R}^2\gtrsim 0.99$ for all fits.
  }
\end{table*}

\begin{figure*}[t!]
  \centering
  \includegraphics[width=0.98\textwidth]{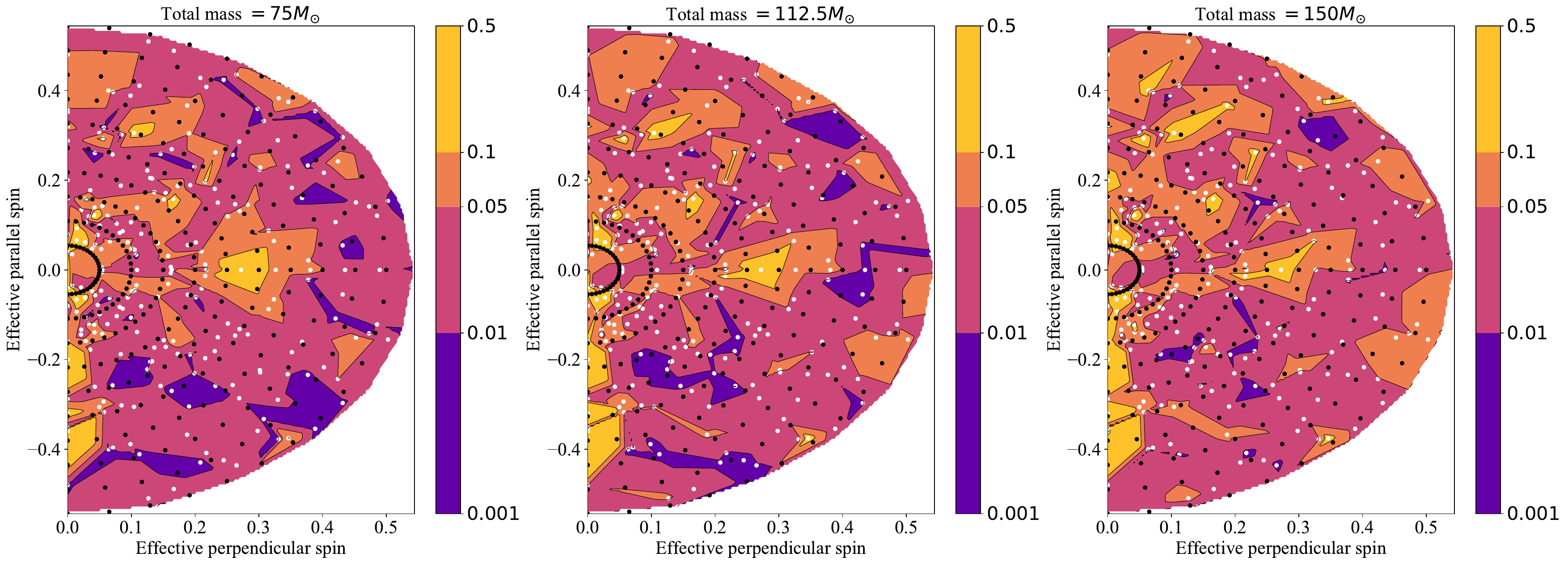}
  \caption{Fit unfaithfulness over the verification set.
    We plot the absolute value of the relative difference, \textit{i.e.}, $\Delta_\text{rel}^k$ [see equation (21)], between the data for the $\log_{10}$ of the
    \NRsur-\modelname{SEOBNRv5PHM} mismatches and the fit (20) to it with equations~(19a, 19b) substituted.
    The panels from left to right correspond to total masses of $75\Msun, 112.5\Msun, 150\Msun$.
    The effective perpendicular, $\chiperp$, and effective parallel spins, $\chipar$, are defined in equations~(11, 12), respectively.
    The white dots mark the verification set data points while the black dots mark the training set points.
    The color scale is logarithmic. Overall, we observe an absolute relative difference of less than $0.05$ for most of the parameter space. }
  \label{fig:verification_contour_plot}
\end{figure*}

\begin{figure}[t!]
  \centering
  \includegraphics[width=0.48\textwidth]{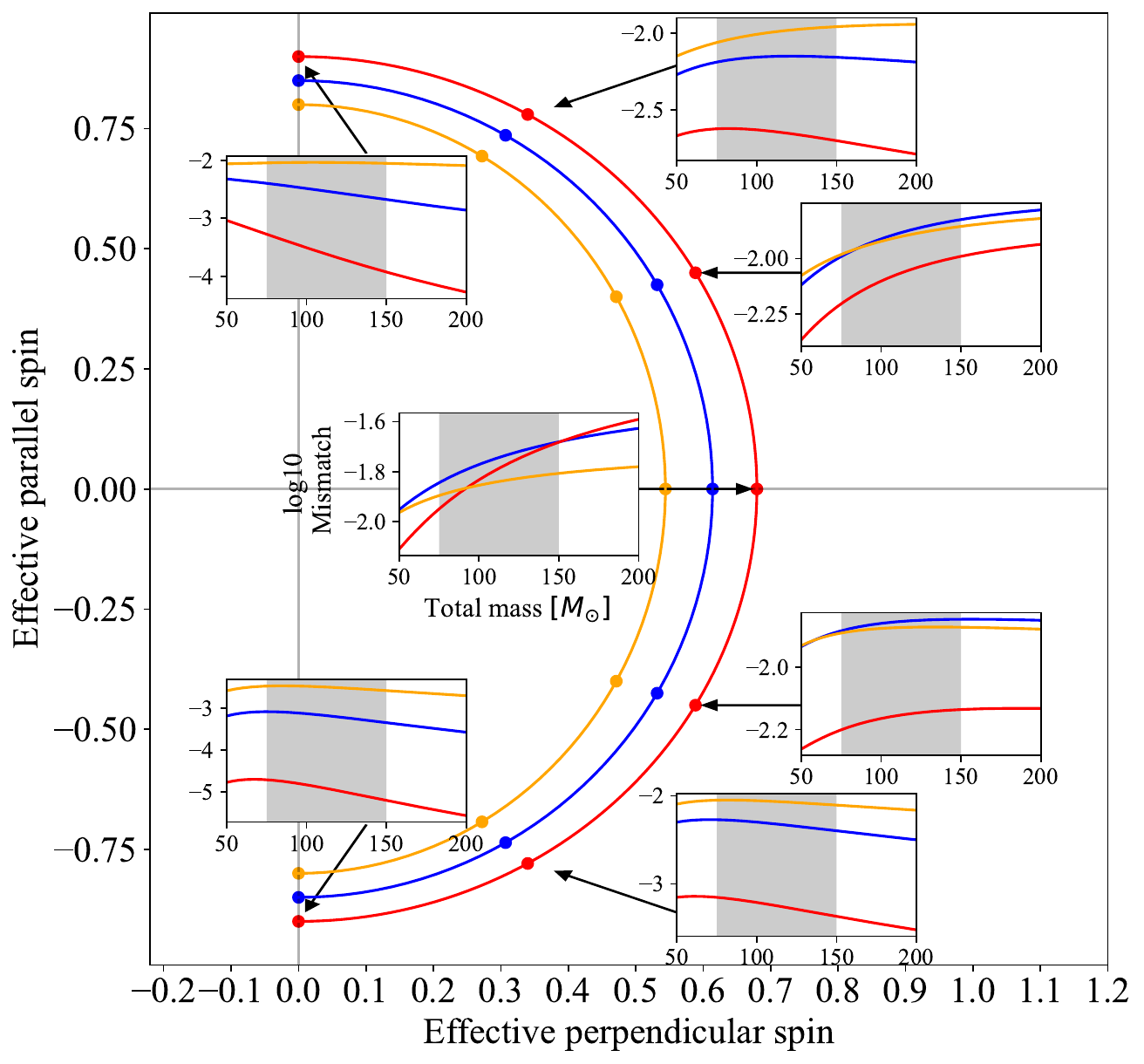}
  \caption{Extrapolation of the fits. We show the performance of the fit to the
    $\log_{10}$ of the \NRsur-\modelname{SEOBNRv5PHM} mismatches extrapolated to
    $\{q,\bar{S}_{1,2}\}=\{1/5,0.85\}$ (blue) and $\{1/6,0.9\}$ (red) as well as
    the fit at the edge of its training region with $\{q,\bar{S}_{1,2}\}=\{1/4,0.8\}$ (orange).
    The red, blue, orange dots accordingly mark the positions of seven cases along each corresponding
    ellipse with the above $\{q,\bar{S}_{1,2}\}$ values and the remaining intrinsic parameters
    chosen such that the dots trace each ellipse in angular steps of $\pi/6$.
    The smaller orange ellipse was shown previously in Supplementary Figure 4.
    Pointing to each $\Phi = \text{constant}$ dot is an inset showing the plot of the fit from $\Mtot=50\Msun$ to $200\Msun$, but at each separate elliptical coordinate.
    The shaded gray region in each inset has been placed to remind the reader our training range
    of $\Mtot\in[75,150]\Msun$.
    From the insets, we see that the extrapolation does not appear to be pathological
    and the blue curves mostly lay between the red and orange ones as expected.}
\end{figure}

\end{document}